\documentclass[12pt]{article}
\usepackage{graphicx,psfrag,epsfig}
\usepackage{amssymb,amsmath,amscd,amsthm}
\usepackage{graphicx,psfrag,epsfig}

\usepackage{graphicx}
\usepackage[active]{srcltx}

\newtheorem{theorem}{Theorem}[section]
\newtheorem{corollary}[theorem]{Corollary}

\newtheorem{lemma}[theorem]{Lemma}
\newtheorem{proposition}[theorem]{Proposition}
\newtheorem{definition}[theorem]{Definition}

\date{}

\begin{document}

\date{}
\title{Bargmann type estimates of the counting function for general Schr\"{o}dinger operators}
\author{
 S.
Molchanov\footnote{Dept of Mathematics, University of North
Carolina, Charlotte, NC 28223, smolchan@uncc.edu},
B. Vainberg\footnote{Dept of Mathematics, University of North
Carolina, Charlotte, NC 28223, brvainbe@uncc.edu ; corresponding author}
}
 \maketitle
\begin{abstract}
The paper concerns upper and lower estimates for the number of negative eigenvalues of one- and two-dimensional Schr\"{o}dinger operators and more general operators with the spectral dimensions $d\leq 2$. The classical Cwikel-Lieb-Rosenblum (CLR) upper estimates require the corresponding Markov process to be transient, and therefore the dimension to be greater than two. We obtain CLR estimates in low dimensions by transforming the underlying recurrent process into a transient one using partial annihilation. As a result, the estimates for the number of negative eigenvalues are not translation invariant and contain Bargmann type terms. The general theorems are illustrated by analysis of several classes of the Schr\"{o}dinger type operators (on the Riemannian manifolds, lattices, fractals, etc.).  We provide estimates from below which prove that the results obtained are sharp. Lieb-Thirring estimates for the low-dimensional Schr\"{o}dinger operators are also studied.
\end{abstract}


\section{Introduction}

Let $N_0(V)=\#\{\lambda_j\leq 0\}$ be the number of non-positive eigenvalues of a Schr\"{o}dinger operator
\begin{equation} \label{sc}
H=-\Delta-V(x),~V\geq 0,
\end{equation}
on $R^d$ or $Z^d$. Everywhere below we assume that the potential is non-negative. The standard approach to Cwikel-Lieb-Rosenblum (CLR) estimates for $N_0(V)$ (see \cite{c}, \cite{L}-\cite{Lt1}, \cite{r}, \cite{rs}) requires the Markov process $x(t)$ which corresponds to the unperturbed operator $H_0=-\Delta$ to be transient. In the lattice case, when $x(t)$ is the symmetric random walk on $Z^d$, the transience means that the expectation of the total time the process $x(t)$ spends in the initial point is finite. The latter is equivalent to the condition
\begin{equation}\label{rec}
\int_0^{\infty}p_0(t,x,x)dt<\infty,
\end{equation}
where $p_0(t,x,y)$ is the fundamental solution of the corresponding parabolic problem
$$
\frac{dp_0}{dt}=\Delta p_0,~t>0,~~ p_0(0,x,y)=\delta_y(x).
$$
In the continuous case (when $x(t)$ is a Brownian motion), one needs to talk about the time spent in a neighborhood of the initial point (not at the point itself), and the transience means that
\begin{equation}\label{rec1}
\int_0^{\infty}\int_{B_{\varepsilon} (x)} p_0(t,x,y)dydt<\infty,
\end{equation}
where $B_\varepsilon (x)=\{y:|x-y|<\varepsilon\}$ is a neighborhood of the point $x$.

Conditions (\ref{rec}),(\ref{rec1}), obviously, do not depend on $x$. The integrals (\ref{rec}),(\ref{rec1}) diverge for recurrent processes. In both lattice and continuous cases $p(t,x,y)\sim \frac{c_0}{t^{d/2}},~t\to\infty, ~|x-y|<C,$ and the process $x(t)$ is recurrent when $d=1,2$ and transient when $d\geq 3$. The CLR estimates are valid for more general operators than (\ref{sc}) (see \cite{2,1,mv} and references there), but usually the transience is an essential requirement when these more general operators are considered.

Recall one of the forms (not the most general) of the CLR estimate for Schr\"{o}dinger-type  operators $H=H_0-V(x)$ on $L^{2}(X,%
\mathcal{B,}\mu )$ where
 $X$ is a complete $\sigma $-compact metric space with the
Borel $\sigma $-algebra $\mathcal{B}(X)$ and a $\sigma $-finite measure $\mu
(dx).$ Let $H_{0}$ be a self-adjoint non-negative operator  such that

(a) the parabolic problem
$$
\frac{\partial p_0}{\partial t}+H_0p_0 =0, ~~t>0, \quad p_0(0,x,y)=\delta_y(x),
$$
has a unique solution in the class of symmetric non-negative probability densities on $(0,\infty)\times X\times X$  (i.e., $\int_Xp_0(t,x,y)\mu(dy)=1$), and

(b) the integral operators
$$
(P_tf)(x)=\int_Xp_0(t,x,y)f(y)\mu(dy)
$$
form a strongly continuous Markov semigroup $P_t $ acting
on $C(X)$. Then there exists a standard Markov process $x(t)$ with the generator $-H_0$ and the transition density $p_0(t,x,y)$. This process has strong Markov property and right continuous trajectories, see \cite{Dynkin}. Obviously, condition (b) holds if function $p_0$ is continuous.

If conditions (a), (b) hold, then
\begin{equation} \label{1xx}
N_0(V) \leq \frac{1}{c(\sigma )}\int_{X}V(x)\int_{\frac{\sigma }{V(x)}%
}^{\infty }p_0(t,x,x)dt\mu (dx),~V\geq 0,
\end{equation}
where $\sigma >0$ is arbitrary and $c(\sigma )=e^{-\sigma }\int_{0}^{\infty }\frac{ze^{-z}dz}{z+\sigma }.$

Formula (\ref{1xx}) is meaningful only if function $p(x)=\int_{1}^{\infty }p_0(t,x,x)dt$ is locally $\mu$-summable. If the underlying Markov process $x(t)$ is recurrent, then $p(x)=\infty$ for all $x$, and (\ref{1xx}) is useless (the right-hand side is infinity). Function $p(x)$ can also be equal to infinity identically or be finite almost $\mu$-everywhere, but not locally summable for transient processes (see examples in section 2). However, it is locally summable for ``typical" transient processes (i.e., under mild assumptions). Thus, in order to apply the estimate  (\ref{1xx}) one needs the process to be transient.

A widespread estimate
\begin{equation}\label{c}
N_0(V)\leq C_{d}\int_{R^{d}}V^{\frac{d}{2}}(x)dx,~V\geq 0,
\end{equation}
for the Schr\"{o}dinger operator (\ref{sc}) in $R^d,~d\geq 3$, follows immediately from (\ref{1xx}) since $p_0(t,x,x)=c_d t^{-d/2}$ in this case. The restriction on the dimension is very essential here. Indeed, the following three facts are valid for the latter operator in dimensions one and two: the process $x(t)$ is recurrent, the integral (\ref{1xx}) diverges (the formula is correct, but useless), and formula (\ref{c}) is not valid. In order to justify the latter fact one needs to note that $N_0(V)\geq 1$ for operator (\ref{sc}) in dimensions $d=1,2$ (if $V\geq 0$ is not identically zero), i.e., the estimate for $N_0(V)$ can't be homogeneous in $V$, see \cite{sim}.

This paper concerns the estimates of $N_0(V)$ (and of the Lieb-Thirring sums $S_\gamma=\sum_{\lambda_j<0}|\lambda_j|^\gamma$) for general operators $H=H_0-V$ on $L^{2}(X,\mathcal{B,}\mu )$ in the case when assumptions (a), (b) hold, but the underlying Markov process $x(t)$ on $X$ with the generator $-H_0$ is recurrent. In particular, new results will be obtained for operators (\ref{sc}) in dimensions $d=1,2$.

The well known Bargmann estimate for 1-D Schr\"{o}dinger operator
$$
H=H_0-V(x)=-\frac{d^2}{dx^2}-V(x),~~V\geq 0,
$$
has the form
\begin{equation} \label{barg}
N_0(V)\leq 1+\int^\infty_{-\infty}|x|V(x)dx.
\end{equation}
We will present an abstract form of the Bargmann and refined Bargmann type estimates and illustrate it with numerous examples.

The main general result can be formulated briefly as follows. Let the process  $x(t)$ be recurrent. Then we introduce a new process using the annihilation (or ``killing") with the rate $q(x)\in C_{com}(X),~q\geq 0$, i.e., the new transition density $p_1$ satisfies
$$
\frac{dp_1}{dt}=-H_0 p_1-qp_1,~t>0,~~ p_0(0,x,y)=\delta_y(x).
$$
\begin{theorem}\label{1.1}
Under weak assumptions on $H_0$ stated in section 3,
the new process with the transition density $p_1$ is transient, and the following inequality holds:
\begin{equation} \label{gint}
N_0(V)\leq 1+\frac{1}{c_1(\sigma )}\int_{X}V(x)\int_{\frac{\sigma }{V(x)}%
}^{\infty }p_1(t,x,x)dt\mu (dx),~V\geq 0,
\end{equation}
when $\|q\|_1=\int qd\mu$ is small enough
\end{theorem}

\textbf{Remark 1.} As it was mentioned above, the transience does not guarantee the convergence of the interior integral in (\ref{gint}) even for continuous compactly supported $q$. We will consider particular classes of operators where
$p_1$ can be estimated, the integral converges and can be expressed more explicitly.

\textbf{Remark 2.}
The Bargmann estimate (\ref{barg}) is an immediate consequence of this general result (\ref{gint}) with $\sigma=0$. The same general result with $\sigma>0$ provides a generalization of the Bargmann estimate for the operator $H_0=-\frac{d^2}{dx^2}$ which is valid for a wider class of potentials (we called this estimate the refined Bargmann estimate).

The following result is one of the consequences of Theorem \ref{1.1}.
\begin{theorem}\label{t1vv}
Let $X$ be a countable set, and let $H_0$ be a symmetric non-negative operator on $L^2(X)$
\begin{equation}\label{genh}
H_0\psi(x)=\sum_{y\in X}h(x,y)\psi(y),
\end{equation}
where
\begin{equation}\label{above11}
h(x,y)\leq 0~~\text{if}~~x\neq y,\quad \sum_{y\in X}h(x,y)= 0 ;\quad h(x,x)\leq c_0 ~~~\text {for all} ~~x\in X,
\end{equation}
and the following connectivity condition holds: $X$ can not be split in two disjoint non-empty sets $X_1\cup X_2$ in such a way that $h(x_1,x_2)=0$ for each $x_1\in X_1,~x_2\in X_2$.

Assume that $H_0$ is left-invariant with respect to some transitive group $\Gamma$ acting on $X$ ($h(gx,gy)=h(x,y),~g\in \Gamma$) and the underlying random process with the transition probability $p_0(t,x,y)$ is recurrent. Put
$$
R_\lambda^{(0)}(x,y)=-\int_0^\infty e^{-\lambda t}p_0(t,x,y)dt,\quad \lambda>0.
$$

Then, for a fixed $x_0\in X$, the limit (a regularized resolvent)
$$
\widetilde{R}^{(0)}(x,x_0)=2\lim_{\lambda\to +0}[R_\lambda^{(0)}(x,x_0)-R_\lambda^{(0)}(x_0,x_0)]
$$
exists and
$$
N_0(V)\leq 1+\sum_{x\in X}V(x)\widetilde{R}^{(0)}(x,x_0).
$$
\end{theorem}

The paper is organized as follows. Section 2 concerns some general properties of the transient and recurrent Markov processes. In section 3 we introduce the method of partial annihilation and prove Theorem \ref{1.1}. A particular case of the annihilation will also be considered in section 3, which concerns the situation when some points $x_0\in X$ have positive capacity and the function $\tau_0=\min(t: x(t)=x_0)$ is defined. In this case, the annihilation approach can be reduced to the rank-one perturbation technique.

In section 4 we apply Theorem \ref{1.1} to a class of Schr\"{o}dinger operators on a model Riemannian manifold $M$ (see \cite{grig}, \cite{grig-salov}) which corresponds, roughly speaking, to a surface of revolution in $R^d$ (the case $M=R^d$ is included). In the simplest case considered in this section, the Hamiltonian $H_0$ has the form
$$
H_0=(1+|x|^2)^{-\alpha/2} \sum_{i=1}^d\frac{\partial}{\partial x_i}(1+|x|^2)^{\alpha/2}\frac{\partial}{\partial x_i}=
\Delta+\frac{\alpha x\cdot\bigtriangledown}{1+|x|^2},\quad x\in R^d, \quad d+\alpha\geq 0.
$$
The kernel $p_0(t,x,y)$ of the semigroup $e^{-tH_0}$ has the following properties \cite{grig}, \cite{grig-salov}:
$$
p_0(t,x,x)\sim \frac{a}{t^{d/2}},~t\to 0,~p_0(t,x,x)\sim \frac{a}{t^{(d+\alpha)/2}},~t\to \infty,
$$
i.e., operator $H_0$ has different local and spectral dimensions (defined by the above asymptotics of $p_0$), and the Markov process $x(t)$ with generator $H_0$ is transient if $d+\alpha>2$ and recurrent if $d+\alpha\leq 2$. By combining Theorem \ref{1.1} and a generalization of Li-Yau inequalities \cite{li}, \cite{GNY} for the heat kernel of the Schr\"{o}dinger operators on Riemannian manifolds proved by Grigorian \cite{grig}, we obtain an estimate for the counting function $N_0(V)$ for operator $H=H_0+V$ in all the cases when the underlying process is transient or recurrent. We will state here this result only in the borderline case $d+\alpha=2$ which includes the classical 2-D Schr\"{o}dinger operator:

\begin{theorem}
If $d+\alpha=2$ and $d\geq 3$ then
$$
N_0(V)\leq 1+ C\left(\int_{\langle x\rangle^{2}V\leq 1}V\frac{\ln^2\langle x\rangle}{\ln\frac{1}{V}}dx+\int_{\langle x\rangle^{2}V>1}\langle x\rangle^{-\alpha}V^{d/2}dx\right), \quad \langle x\rangle=2+|x|.
$$

If $d=2,~\alpha=0$ (i.e., $H$ is the classical 2-D Schr\"{o}dinger operator) then
\begin{equation}\label{rr}
N_0(V)\leq 1+ C\left(\int_{\langle x\rangle^{2}V\leq 1}V\frac{\ln^2\langle x\rangle}{\ln\frac{1}{V}}dx+\int_{\langle x\rangle^{2}V>1}V\ln(\langle x\rangle^{3 }V)dx\right),
\end{equation}

or, in a bit roughened form,
\begin{equation}\label{rn2aa}
N_0(V)\leq 1+ C\left(\int_{R^2}V(x)\ln\langle x\rangle dx+\int_{V>1}V(x)\ln V(x)dx\right).
\end{equation}
\end{theorem}
Let us note that the right scaling $N_0(\alpha V)\sim \alpha^{d/2},~\alpha\to\infty$ (which agrees with the quasi-classical asymptotics) is valid when $d>3$ and depends on the local dimension of the operator. If $d=2$, then the scaling $N_0(\alpha V)\sim \alpha\ln\alpha,~\alpha\to\infty,$ is logarithmically weaker than expected from the quasi-classics.

In section 5, we continue the discussion of the two-dimensional Schr\"{o}dinger operator (\ref{sc}). One can find some references in section 5. Here we would like to mention only two results. M. Solomyak \cite{solO} obtained an estimate on $N_0(V)$ which is in agreement with quasi-classical asymptotics ($N_0(\alpha V)=O(\alpha),~\alpha\to\infty$). The estimate is rather complicated and expressed in terms of local Orlich norms of the potential. There is a very elegant conjecture \cite{many} that, in the case of the $2$-D Schr\"{o}dinger operator,
\begin{equation}\label{conm1}
N_0( V)\leq 1+C_1\int_{r<r_0}V^\ast(r)\ln \frac{r_0}{r}dx+C_2\int_{r>r_0}V(x)\ln\frac{r}{r_0} dx+C_3\int_{R^2}V(x)dx,\quad r=|x|,
\end{equation}
with arbitrary $r_0>0$, some specific constants $C_j$ and with $V^\ast$ being the monotone spherical rearrangement of $V$. One of the consequences of our result (\ref{rn2aa}) is a justification of (\ref{conm1}) when $\|V\|_{L_1}\leq 1$ or $V$ has a compact support of a fixed size. In the latter case the statement is as follows
\begin{theorem}\label{td22}
Let $V(x)=0$ for $|x|>r_0$. Then
$$
N_0(V)\leq  1
+C\int_{|x|<r_0}V^\ast(r)\ln\frac{2r_0}{ r}dx.
$$
\end{theorem}
Note that this result has the right quasi-classical scaling.

Section 6 is devoted to the application of the general result from section 3 to the case of two-dimensional lattice Schr\"{o}dinger operator. We can't use a powerful differential geometry technique \cite{grig}, \cite{grig-salov} in a non-smooth case of $Z^d$, and the needed estimates of the heat kernel will be obtained here by a direct analytical method. We provide here two results from section 6.
\begin{theorem}
The following estimate holds for the lattice two-dimensional Schr\"{o}dinger operator (\ref{sc}).
\begin{equation}\label{234}
N_0( V)\leq 1+C\sum_{x\in Z^2}V(x)\ln\langle x\rangle dx.
\end{equation}
\end{theorem}
The corresponding theorem in section 6 contains also the refined Bargmann estimate for the same operator. Note that (\ref{234}) has the right quasi-classical order in $V$.

The second result in section 6 concerns the first time $\tau_x$ when the random walk on $Z^2$ starting at point $x$ visits the origin.
\begin{theorem}
The following relation holds for each $\alpha\in R$:
$$
P_x\{\frac{\ln\tau_x}{\ln|x|}\leq \alpha\}\rightarrow \frac{(\alpha-2)_+}{\alpha} \quad \text{as} \quad |x|\to \infty.
$$
Here $(\alpha-2)_+=\max(0,\alpha-2)$.
\end{theorem}

Section 7 contains the general Bargmann type estimates in the case of discrete recurrent Markov chains with the generator $H_0$ defined by (\ref{genh}), (\ref{above11}). An analogue of Theorem \ref{t1vv} is proved there without the assumption on the group symmetry. It is also proved there that in the case of the recurrent underlying Markov chain, for an arbitrary positive sequence $\{a_n\}$, there exists a sequence $\{x_n\}$ for which the operator $H=H_0-V$ with the potential $V=\sum_0^\infty a_n\delta_{x_n}(x)$ has infinitely many negative eigenvalues. Thus, the space invariant estimate of the form (\ref{c}) can not be valid in the recurrent case. Similar result for the standard lattice Schr\"{o}dinger operator in dimensions one and two can be found in \cite{sim}, \cite{birL}, \cite{many}.

A short section 8 contains estimates on $N_0$ from below. A typical result is as follows:
\begin{theorem}
Consider the operator $H=-\Delta-V(x),~V\geq0$ on $Z^2$. If $\sum_{x\in Z^2}V(x)=\infty$, then $N_0(V)=\infty$.
\end{theorem}
A similar fact in the continuous case follows from a much more profound estimate by Grigorian-Netrusov-Yau \cite{GNY}. Their methods do not work in the discrete case.

Some results on the fractal degrees of the 1-D Schr\"{o}dinger operator $(-\Delta)^\alpha,~0<\alpha<1,$ are presented in section 9. The spectral dimension of this operator is $1/\alpha\in(1,\infty)$, and the corresponding random walk is recurrent for $\alpha\geq 1/2$. For example, the following result for $\alpha=1/2$, similar to the one for the 2-D Schr\"{o}dinger operator, will be proved.
\begin{theorem}
If $H_0=(-\Delta)^{1/2}$ on $Z^1$, then
$$
N_0(V)\leq 1+\sum_{x\in Z}\min (1,~C V(x)\ln(2+|x|)).
$$
\end{theorem}

Estimates on the Lieb-Thirring sums $S_\gamma=\sum_{\lambda_j\leq 0}|\lambda_j|^\gamma$ for recurrent underlying processes are discussed in section 10. The classical approach \cite{Lt}, \cite{Lt1} requires the inequality $\frac{d}{2}+\gamma>1$, where $d$ is the spectral dimension. The estimates with exact constants in the important one-dimensional borderline case $d=1,~\gamma=1/2$, were obtained in \cite{dirk}, \cite{hs}. The annihilation method leads to an estimate on $S_\gamma$ for arbitrary $d$ and $\gamma>0$. This estimate is weaker than the one in  \cite{dirk}, \cite{hs}. However, it covers the case $\frac{d}{2}+\gamma<1$. Besides, it does not contain the singular factor $(\frac{d}{2}+\gamma-1)^{-1}$ which appears in the Lieb-Thirring estimate when $\frac{d}{2}+\gamma\downarrow 1$, and therefore it is better than the classical estimate when $\frac{d}{2}+\gamma- 1>0$ is small enough.

Section 11 is devoted to the spectral analysis of the Schr\"{o}dinger operator on a simple fractal, the so-called Dyson's hierarchical lattice. The spectral dimension $s>0$ for this operator can be an arbitrary positive number.  The underlying Markov chain (hierarchical random walk) is transient if $s>2$ and recurrent if $s\leq 2$.

The CLR estimate is applicable in the transient case $s>2$ and contains  the factor $(s-2)^{-1}$ which explodes as $s\downarrow  2$. Our approach with annihilation provides an estimate for $N_0(V)$ which is valid in both cases $s\gtrless 2$ and with a uniformly bounded constant. In particular, it improves the CLR estimate when $s>2$ is close to $2$. Note that this is a common feature for all the problems in sections 9-11 where the spectral dimension changes continuously.

The authors are grateful to O. Safronov for productive discussions and to B. Simon for useful critical remarks.

\section {Transient and recurrent Markov processes.}
The goal of this section is to recall (derive) some properties of the general Markov processes $x(t)$ needed for better understanding of the results of the following sections. We will consider the processes with the generator $-H_0$ on $L^2(X,\mathcal{B},\mu )$ for which conditions (a), (b) hold, i.e., the processes which have strong Markov property and right-continuous trajectories. Moreover, in this section we impose two additional conditions. We assume:

(b$'$) Function $p_0$ is continuous. Recall that this implies assumption (b). All the arguments below remain valid if we replace (b$'$) by the requirement that the semigroup $P_t$ maps bounded measurable functions into continuous ones. This assumption is stronger than condition (b), but weaker than (b$'$).

We also assume the ``strong connectivity":

(c) For any two compacts $K_1,K_2\in \mathcal{B}$, one can find $t_0=t_0(K_1,K_2)>0$ and $\varepsilon=\varepsilon(K_1,K_2)>0$ such that
\begin{equation}\label{scon}
p_0(t_0,x,y)\geq \varepsilon,\quad x\in K_1,~~y\in K_2.
\end{equation}

The standard CLR estimates are meaningful only if the process $x(t)$ is transient. Let us recall the definition of transience. If the metric space $X$ is a countable set and the Markov chain $x(t),~t\geq 0,$ with the generator $-H_0$ (time is continuous) is connected (i.e., $p_0(t,x,y)>0$ for any $t>0,~x,y\in X,$) then the following dichotomy is well known: either
$$
\int_0^\infty p_0(t,x,x)dt<\infty \quad \text{for each}~x\in X
$$
(these chains are called transient) or
$$
\int_0^\infty p_0(t,x,x)dt=\infty \quad \text{for each}~x\in X
$$
(recurrent chains). It is also well known that the transient chains spend ($P$-a.s.) a finite time in each state:
$$
T_{x,y}=\int_0^\infty \delta_y(x(s))ds<\infty,~~x(0)=x,
$$
and the integral above is $P$-a.s. infinite for each $x,y\in X$ if a chain is recurrent.

There are many different definitions of the transient and recurrent processes for the general (continuous) space $X$. In our case (conditions (a) and (b$'$) hold), the most natural definition is the following.
\begin{definition}\label{def}
The process $x(t)$ is called transient if
$$
E(x,K):=E_x\int_0^\infty I_K (x(s))ds =\int_0^\infty dt\int_K p_0(t,x,y)\mu(dy)<\infty
$$
for each $K\in \mathcal{B}, ~0<\mu(K)<\infty,$ and almost every $x\in X$. The process is called recurrent if the latter double integral is infinite for each $K\in \mathcal{B}, ~0<\mu(K)<\infty,$ and a.e. $x\in X$.
\end{definition}

The following dichotomy is similar to the one in the discrete case:
\begin{proposition}\label{pr1}
If conditions (a)-(c) hold, then either $E(x,K)<\infty$ for each $K\in \mathcal{B}, ~0<\mu(K)<\infty,$ and a.e. $x$ (the process $x(t)$ is transient) or $E(x,K)=\infty$ for each $K\in \mathcal{B}, ~0<\mu(K)<\infty,$ and each $x\in X$ (the process $x(t)$ is recurrent).
\end{proposition}
\textbf{Proof.} Consider two compacts $K_1,K_2\subset \mathcal{B}$, and let $t_0'=t_0(K_2,K_1)$ (the order of compacts is reversed). Then, for $x,y \in K_1$, we have
\begin{equation*}
p_0(t+t_0+t_0',x,y)\geq \int_{K_2}p_0(t_0,x,z)p_0(t,z,z')\mu(dz)\int_{K_2}p_0(t_0',z',y)\mu(dz'),
\end{equation*}
and after the integration, this implies that
\begin{equation}\label{rectr}
\int_{K_1}\int_{K_1}p_0(t+t_0+t_0',x,y)\mu(dx)\mu(dy)\geq\varepsilon(K_1,K_2)\varepsilon(K_2,K_1)
\int_{K_2}\int_{K_2}p_0(t,z,z')\mu(dz)\mu(dz').
\end{equation}

Since $\int_Xp_0(t,x,y)\mu(dy)=1$, it follows that
$$
\int_0^Tdt\int_{K}\int_{K}p_0(t,x,y)\mu(dx)\mu(dy)\leq T\mu(K),
$$
and therefore (\ref{rectr}) implies that either
\begin{equation}\label{dix}
\int_0^\infty dt\int_{K}\int_{K}p_0(t,x,y)\mu(dx)\mu(dy)=\infty
\end{equation}
for each $K\in \mathcal{B}, ~0<\mu(K)<\infty,$ or this integral is finite for each $K$.

Similarly, for any $x\in X$ and compact $K$,
\begin{eqnarray*}
\int_{K}p_0(t,x,y)\mu(dy)&\geq& \int_{K}\int_{K}p_0(t_0,x,z)p_0(t-t_0,z,y)\mu(dz)\mu(dy)\\&\geq& \varepsilon(x.K) \int_{K}p_0(t-t_0,z,y)\mu(dz)\mu(dy), \quad t_0=t_0(x,K),
\end{eqnarray*}
which implies that
\begin{equation}\label{dix1}
t_0\mu(K)+\int_0^\infty dt\int_{K}p_0(t,x,y)\mu(dy)\geq \varepsilon(x.K)\int_0^\infty dt\int_{K}\int_{K}p_0(t,x,y)\mu(dy)\mu(dz).
\end{equation}

If (\ref{dix}) holds for each $K$, then (\ref{dix1}) implies that $E(x,K)=\infty$ for each $x\in X$ and $K$, i.e., the process is recurrent. If integral (\ref{dix}) is finite for each $K$, then $E(x,K)<\infty$ for a.e. $x\in X$ and each $K$ due to the Fubini theorem, i.e., the process is transient.
\qed

The definitions of the transient and recurrent processes were introduced based on whether the expectation of the time spent in each compact is finite or infinite. The next proposition shows that this definition agrees with the one when the time (and not the expectation) is measured. Let us denote by $T_{x,K}$ the total time which process $x(t)$ starting at a point $x\in X$ spends in a compact $K$.
\begin{proposition}\label{pr2}
Let conditions (a), (b$'$), (c) hold and $0<\mu(K)<\infty$. Then $T_{x,K}<\infty$ $P$-a.s. for each $K$ and a.e. $x\in X$ if the process is transient, and $T_{x,K}=\infty$ $P$-a.s. for each $K$ and $x\in X$ if the process is recurrent.
\end{proposition}
\textbf{Proof.} The statement concerning the transient processes is obvious (if the expectation of a random variable is finite, then the variable is finite $P$-a.s.). The second part of the proposition is more delicate, and its proof is based on a construction of an imbedded (or ``loop") Markov chain. Due to the continuity of $p_0$ and the connectivity condition (c), it is enough to prove the statement when $K$ is a fixed ball $B$. It will be done for a ball $B_{1}$ of a fixed radius $r>0$
 centered at some point $x_0\in X$ where $r$ is large enough so that there exists a smaller ball $B_0$ centered at the same point and located strictly inside of $B_1$.

Due to the connectivity, there exist $t=t_0$ and $\gamma>0$ such that $\int_{X\setminus B_{1}}p_0(t_0,x,y)\mu(dy)>\gamma>0,~~x\in B_0$. Let us construct the following sequence of Markov moments $\tau_n, \theta_n$ related to transitions of the process $x(t)$ from $B_0$ to $X\setminus B_{1}$ and back:
$$
\tau_1=\min (t: x(t)\in X\setminus B_{1}),
$$
$$
\theta_1=\min (t\geq \tau_1: x(t)\in B_{0}),
$$
$$
\tau_2=\min (t\geq \theta_1: x(t)\in X\setminus B_{1}), \quad \text{etc}.
$$
The connectivity condition implies
$$
P_x\{\tau_1>t_0\}\leq (1-\gamma)<1,\quad \text{for all}~~ x\in B_0,
$$
where $P_x\{A\}$ is the probability of the event $A$ for the random process starting at point $x$. By induction,
$$
P_x\{\tau_1>nt_0\}\leq (1-\gamma)^n<1,\quad \text{for all}~~ x\in B_0.
$$
Together with the Feller property this implies that
$$
0<c_1<E_x\tau_1<c_2<\infty.
$$
Consider now $\theta_1$. Due to the connectivity and strong Markov and Feller properties of the process, there are two possibilities:
\begin{equation*}
P_x\{\theta_1<\infty\}<1-\delta<1,~~~x\in B_0, ~~\text{or} ~~P_x\{\theta_1<\infty\}=1,~~~x\in B_0.
\end{equation*}
In the first case, there are $P$-a.s. only finitely many loops between $B_0$ and $X\setminus B_{1}$, and this implies that the process is transient. The number of the loops is $P$-a.s. infinite in the second case. This leads to the recurrency of the process. Indeed, one can consider the ergodic Markov chain $z_n=x(\theta_n)$ on the compact $B_0$. The mean transition time from $B_0$ to $X\setminus B_{1}$ equals $\int_{B_0}E_x(\tau_1)\mu(dx)$, where $\mu$ is the invariant measure of the chain $z_n$. The law of large numbers implies that the time which the process spends in $B_{1}$ grows at least linearly with the growth of the number of loops. Thus $T_{x,B_{1}}=\infty$.
\qed

Let us discuss now the relation between the  transience/recurrence of the processes and the convergence/divergence of the integral
\begin{equation}\label{idio}
I(x)=\int_1^\infty p_0(t,x,x)dt.
\end{equation}
Let us stress that the lower limit here is one since typically the function $p_0(t,x,x)$ is not integrable at $t=0$ in the continuous case.
\begin{proposition}\label{pr3}
Let conditions (a),(b) hold. If $I(x)<\infty$ for a.e. $x\in X$ and $I(x)$ is locally $\mu$-summable, then the process $x(t)$ is transient.
\end{proposition}
\textbf{Proof.} The statement follows immediately from the following inequality.
$$
p_0(t,x,y)=\int_{X}p_0(\frac{t}{2},x,z)p_0(\frac{t}{2},z,y)\mu(dz)\leq \sqrt{\int_{X}p_0^2(\frac{t}{2},x,z)dz}
\sqrt{\int_{X}p_0^2(\frac{t}{2},z,y)dz}
$$
$$
=\sqrt{\int_{X}p_0(\frac{t}{2},x,z)p_0(\frac{t}{2},z,x)dz}
\sqrt{\int_{X}p_0(\frac{t}{2},z,y)p_0(\frac{t}{2},y,z)dz}
$$
$$
=\sqrt{p_0(t,x,x)p_0(t,y,y)}\leq \frac{1}{2}[p_0(t,x,x)+p_0(t,y,y)].
$$
\qed
\begin{proposition}\label{pr4}
Let conditions (a)-(c) hold. If the process $x(t)$ is recurrent, then $I(x)=\infty$ for each $x\in X$.
\end{proposition}
\textbf{Proof.} Let us fix an arbitrary compact $K\subset X$. Let $t_0,\varepsilon$ be the constants defined in condition (c) for $K_1=x, ~K_2=K$. Then
\begin{eqnarray*}
\int_1^\infty p_0(t,x,x)dt &\geq& \int_{2t_0+1}^\infty dt\int_{K}\int_{K}p_0(t_0,x,z_1)p_0(t-2t_0,z_1,z_2)p_0(t_0,z_2,x)\mu(dz_1)\mu(dz_2)\\
&\geq& \varepsilon^2\int_{2t_0+1}^\infty\int_{K}\int_{K}p_0(t-2t_0,z_1,z_2) \mu(dz_1)\mu(dz_2).
\end{eqnarray*}
It was shown in the proof of Proposition \ref{pr1} that (\ref{dix}) holds for recurrent processes. This completes the proof since the right hand side above differs from the integral (\ref{dix}) by at most $\varepsilon^2\mu(K).$
\qed

We conclude this section with two important examples.

\textbf{Example 1.} This example shows that $p(t,x,x)$ can be equal to infinity identically for a transient process. Let time $t=0,1,2,...$ be discrete and
$$
x(t)=x+X_1+X_2+...+X_t, \quad x\in R^d,
$$
where $\{X_n, ~n\geq 1,\}$ are i.i.d.r.v. with density $p(x)$. Then the transition density $p_0(t,x,y)$ of $x(t)$ is given by
$$
p_0(t,x,y)=p_t(x-y)=(p\ast p \ast...\ast p)(x-y),
$$
where the convolution of $t$ factors is taken in the right-hand side above. Let the density $p(x)$ have a strong singularity at $x=0$:
\begin{equation}\label{ppp}
p(x)=c\frac{I_{|x|\leq 1/2}}{|x|^d\ln^2\frac{1}{|x|}}.
\end{equation}
Then
$$
p_n(x)\sim \frac{c_n}{|x|^d\ln^{2n}\frac{1}{|x|}}, \quad x\to 0,
$$
i.e., $p(t,x,x)\equiv \infty,~~t\geq 1.$ At the same time, the process $x(t)$ is transient if $d\geq 3$.

In order to justify the latter statement one needs to show (see Definition \ref{def}) that, for each $a>0$,
$$
\sum_{n=1}^\infty \int_{|y|<a}p_n(x-y)dy<C(a)<\infty , \quad d\geq 3.
$$
Hence, it is enough to prove that
$$
\sum_{n=1}^\infty \int_{R^d}e^{-|y|^2}p_n(x-y)dy<\infty , \quad d\geq 3,
$$
which is equivalent to
$$
\sum_{n=1}^\infty \int_{R^d}e^{-|k|^2}\widehat{p}_n(k)dk<\infty , \quad d\geq 3,
$$
where $\widehat{p}_n$ is the Fourier transform of $p_n$. Since $\widehat{p}_n=(\widehat{p})^n$, it remains to show that
$$
\int_{R^d}e^{-|k|^2}\frac{\widehat{p}(k)}{1-\widehat{p}(k)}dk<\infty , \quad d\geq 3.
$$
Since $p(x)$ is a density of a probability measure and it is even, it follows that $\widehat{p}(k)$ vanishes at infinity, $\widehat{p}(k)<1$ when $k\neq 0$, and $1-\widehat{p}(k)\sim |k|^2$ as $k\to 0$. This implies the convergence of the above integral and the transience of the process $x(t)$ when $d\geq 3$. (The same arguments show that the process is recurrent if $d=1,2$)

\textbf{Example 2.} This example shows that $p(t,x,x)$ can be equal to infinity identically for a transient process with continuous time. Consider the process $x(t),~t\in[0,\infty),$ in $R^d$ with independent increments. The Fourier transform $\widehat{p}_0(t,k)$ of its transition density $p(t,x-y)$ has the form (see \cite{F})
$$
\widehat{p}_0(t,k)=e^{-t\Phi(k)},\quad \text{where}~~\Phi(k)=\int_{R^d}\frac{e^{ikx}-1-i\sin(kx)}{|x|^2}M(dx).
$$
Here $M(dx)$ is a finite measure for $|x|\leq 1$ and $\int_{|x|\geq 1}\frac{M(dx)}{|x|^2}<\infty$. Let the measure $M(dx)$ have a strong singularity at $x=0$, say $M(dx)=p(x)dx$ where $p(x)$ is given by (\ref{ppp}). Then $\Phi(k)$ decreases slowly as $|k|\to\infty$, and
$$
p(t,x-y)|_{y=x}=\frac{1}{(2\pi)^d}\int_{R^d}e^{-t\Phi(k)}dk\equiv \infty.
$$
The transience of the process $x(t)$ when $d\geq 3$ can be shown in the same way as in the previous example.

\section {Two general theorems on annihilation.}
Recall that
$H=H_0-V(x)$ is an operator on $L^{2}(X,\mathcal{B,}\mu )$ (see formula (\ref{1xx})), where
$H_{0}$ is a self-adjoint non-negative operator  such that conditions (a), (b) hold. Thus the operator $-H_0$ is the generator of a strongly continuous Markov semigroup $P_t $ acting
on $C(X)$. The kernel $p_0(t,x,y)$ of $P_t $ is the transition density of the underlying Markov process $x(t).$ We do not make any assumptions on the transience or recurrency of the process $x(t).$

Let us introduce a killing compactly supported potential $q(x)\geq 0$. Let it be continuous (for simplicity), and let $Q\subseteq X$ be its support. Put $H_1=H_0+q(x)$, and denote by $p_1(t,x,y)$ the kernel of the semigroup $e^{-tH_1}$, i.e.,
\begin{equation}\label{gta}
\frac{\partial p_1}{\partial t}=-H_1p_1, \quad p_1(0,x,y)=\delta_y(x).
\end{equation}

Denote by
$$
\widehat{R}^{(0)}_\lambda=\chi R^{(0)}_\lambda\chi, \quad R^{(0)}_\lambda =(-H_0-\lambda)^{-1}, ~~\lambda>0,
$$
the truncated resolvent of the operator $-H_0$. Here $\chi=I_Q$ is the characteristic function of $Q$. Since  $H_{0}\leq 0$, operators $R^{(0)}_\lambda, ~\widehat{R}^{(0)}_\lambda:~L^{2}(X,\mathcal{B,}\mu) \to L^{2}(X,\mathcal{B,}\mu )$ are analytic in $\lambda$ when $\lambda >0.$ Properties of the truncated resolvent $\widehat{R}^{(0)}_\lambda$ at $\lambda=0$ (and $\lambda<0$) for differential operators $H_0$ can be found in \cite{vain75}.
\begin{theorem}\label{nt1}
I. If conditions (a), ($b'$), (c) hold, then the Markov process $\widetilde{x}(t)$ with the generator $-H_1$ is transient, i.e.,
$$
\int_0^\infty \int_Kp_1(t,x,y)dt\mu(dy)<\infty
$$
for each compact $K \in \mathcal{B},~0<\mu(K)<\infty,$ and a.e. $x\in X$ (in fact, the latter inequality holds for all $x \in X$).

II. If conditions (a), (b) hold, then the following estimate is valid for the number $N_0(V)$ of non-positive eigenvalues of the operator $H=H_0-V$:
\begin{equation}\label{gt}
N_0(V)\leq n_0+\frac{2}{c(2\sigma )}\int_{X}V(x)\int_{\frac{\sigma }{V(x)}%
}^{\infty }p_1(t,x,x)dt\mu (dx),~V\geq 0,
\end{equation}
where $c(\sigma)$ is defined in (\ref{1xx}) and $n_0=N_0(q)$.

The constant $n_0$ in (\ref{gt}) is equal to one ($n_0=1$) if {\rm max}$q(x)$ is small enough and the truncated resolvent satisfies the following conditions: $\widehat{R}^{(0)}_\lambda=f(\lambda)T_0+T(\lambda),$ where $T_0$ is a one-dimensional symmetric operator, function $f$ is real, $f(\lambda),|f'(\lambda)|\to \infty$ as $\lambda \to +0$, and $\|T(\lambda)\|<C,\frac{\|T'(\lambda)\|}{f'(\lambda)}\to 0$ as $\lambda \to +0$.
\end{theorem}
\textbf{Remark.} Conditions (a), (b) imply both estimate (\ref{gt}) and the standard CLR estimate (\ref{1xx}). If the underlying process $x_0(t)$ is recurrent, then the estimate (\ref{1xx}) is meaningless (the right-hand side is infinite in this case), while estimate  (\ref{gt}) provides usually a meaningful result (due to the first statement of the theorem). However, we would like to recall that the transience does not guarantee the convergence of the integrals in the right-hand sides of  (\ref{gt}) or (\ref{1xx}). Counterexamples  were given in the end of the previous section.

\textit{Hence, an application of Theorem \ref{nt1} must be accompanied by an estimate on $p_1(t,x,x)$ which justifies the convergence of the integral  (\ref{gt}) and specifies the estimate (\ref{gt}).} This part of the work can be very non-trivial. Many examples will be considered in the following sections.

{\bf Proof.}
Markov process $\widetilde{x}(t)$ is a subprocess of $x(t)$ (see \cite{Dynkin}), i.e., $\widetilde{x}(t)=x(t),~t<\tau,$ where $\tau$ is the annihilation (random) moment for $\widetilde{x}(t)$. While the process $x(t)$ is defined on the probability space $(\Omega,\mathcal{F},P_x)$, the process $\widetilde{x}(t)$ can be realized on $\Omega\times [0,\infty)$ where the measure $\pi_\omega(ds)$ on $[0,\infty)$ is given by the formula
$$
\pi_\omega(ds)=q(x(s))e^{-\int_0^s q(x(u))du}ds.
$$
In other words, the distribution of the random variable $\tau$ is given by
\begin{equation}\label{2345}
\pi_\omega(\tau>t)=e^{-\int_0^tq(x(u))du}.
\end{equation}

Consider a compact $K_0$ such that $0<\mu(K_0)<\infty$ and $q(x)\geq \varepsilon_0>0$ when $x\in K_0$, i.e., the rate of the annihilation on $K_0$ is at least $\varepsilon_0$. From (\ref{2345}) it follows that the mean time that the process  $\widetilde{x}(t)$ (with an arbitrary initial point) spends in $K_0$ does not exceed $1/\varepsilon_0$. Hence, for each $x\in X$,
$$
\int_0^\infty \int_{K_0}p_1(t,x,y)dt\mu(dy)<1/\varepsilon_0.
$$
This implies the transitivity of $\widetilde{x}(t)$ due to Proposition \ref{pr1}. The first statement of  the theorem is proved.

In order to prove the second statement, we write $H$ in the form
$$
H=H_0-V=H_0+q-(q+V)=H_1-(q+V).
$$
The Birman-Schwinger principle implies that
$$
N_0(V)=N_0(q+V;H_1)\leq N_0(2q;H_1)+N_0(2V;H_1),~~H_1=H_0+q(x),
$$
where the second argument of the function $N_0$ is the unperturbed operator. Since $H_1-2q=H_0-q$, the latter inequality implies that
$$
N_0(V)\leq n_0+N_0(2V;H_1).
$$

One obtains inequality  (\ref{gt}) by applying the standard CLR estimate (\ref{1xx}) with $\sigma$ replaced by $2\sigma$ to the second term on the right-hand side above.

It remains to prove the last part (concerning $n_0$) of the second statement of the theorem. We need to show that the eigenvalue problem
\begin{equation}\label{ree}
(H_0-\varepsilon q(x)-\nu)\psi=0, \quad \nu\leq 0, \quad \psi\in L^{2}(X,\mathcal{B,}\mu ),
\end{equation}
has at most one eigenvalue $\nu=\nu(\varepsilon)\leq 0$, and it is simple, provided that $q$ is fixed and $\varepsilon>0$ is small enough. Problem (\ref{ree}) can be rewritten in the form
\begin{equation}\label{re}
\psi+\varepsilon R^{(0)}_\lambda q(x)\psi=0, \quad \lambda=-\nu\geq 0.
\end{equation}
We note that (\ref{re}) is equivalent to $\varphi+\varepsilon \sqrt{q} R^{(0)}_\lambda \sqrt{q}\varphi=0, ~ \lambda\geq 0$ (where $\varphi=\sqrt{q}\psi$).
One can replace here $R^{(0)}_\lambda$ by $\widehat{R}^{(0)}_\lambda$, and this leads to the equation
\begin{equation}\label{re1}
(I+\varepsilon f(\lambda) \sqrt{q}T_0\sqrt{q}+\varepsilon \sqrt{q}T(\lambda)\sqrt{q})\varphi=0, \quad \lambda\geq 0.
\end{equation}

Equation  (\ref{re}) implies that (\ref{re1}) may have a non-trivial solution only for small values of $\lambda$ if $\varepsilon$ is small enough. Thus, we will assume below that both $\lambda$ and $\varepsilon$ are small.

We may assume that $\sqrt{q}T_0\sqrt{q}\neq 0$ since otherwise equation (\ref{re1}) has only trivial solution $\varphi=0$ when $\lambda$ and $\varepsilon$ are small. Hence, there exist a function $\alpha\in L^{2}(X,\mathcal{B,}\mu ),~\|\alpha\|=1,$ and a real constant $\beta\neq 0$ such that $\sqrt{q}T_0\sqrt{q}=\beta P_\alpha$, where $P_\alpha$ is the orthogonal projection (in $L^{2}(X,\mathcal{B,}\mu )$) on $\alpha$. Let us denote by $ P_\alpha^\perp$ the orthogonal projection on the orthogonal complement to $\alpha$.

We represent $\varphi$ as $\varphi=c\alpha+\alpha^\perp$, where $\alpha^\perp=P_\alpha^\perp\varphi$, and we write (\ref{re1}) as a system of two equations for $c$ and $\alpha^\perp$ by applying operators $P_\alpha, ~P_\alpha^\perp$ to both sides of (\ref{re1}). From the assumption on the norm of the operator $T(\lambda)$ it follows that the second equation can be immediately solved for $\alpha^\perp$ when $\lambda,\varepsilon$ are small enough:
$$
\alpha^\perp=-c[1+\varepsilon P_\alpha^\perp\sqrt{q}T(\lambda)\sqrt{q}]^{-1}(\varepsilon P_\alpha^\perp\sqrt{q}T(\lambda)(\sqrt{q}\alpha)), \quad \lambda\geq 0,
$$
and the solution has the form
\begin{equation}\label{al}
\alpha^\perp=\varepsilon c h (\lambda,\varepsilon; x), \quad {\rm where} ~~\|h\|<C,~\frac{\|\frac{d}{d\lambda}h\|}{f'(\lambda)}\to 0, \quad {\rm as}~~ \lambda+|\varepsilon|\to  0.
\end{equation}
The latter estimates for $\|h\|,\|\frac{d}{d\lambda}h\|$ hold for small $\lambda,\varepsilon$, and they follow from the assumption on the norm of the operator $T$.

Applying $P_\alpha$ to (\ref{re1}) and using (\ref{al}), we arrive at the following equation for $c$:

\begin{equation}\label{ccc}
[1+\varepsilon\beta  f (\lambda) +\varepsilon\gamma(\lambda,\varepsilon)]c=0, \quad \gamma=\langle\sqrt{q}T(\lambda)\sqrt{q}(\alpha+\varepsilon h),\alpha \rangle, \quad \lambda\geq 0.
\end{equation}
Here $\gamma$ satisfies the estimates similar to estimates (\ref{al}) for $h$, i.e.,
\begin{equation}\label{gam}
|\gamma|<C,~\frac{|\frac{d}{d\lambda}\gamma|}{f'(\lambda)}\to 0 \quad {\rm as}~~ \lambda+|\varepsilon|\to  0.
\end{equation}
Since $\varphi$ is uniquely defined by $c$, it remains to show that the equation
$$
F(\lambda,\varepsilon):=1+\varepsilon \beta f(\lambda)+\varepsilon\gamma (\lambda,\varepsilon)=0, ~\lambda\geq 0,
$$
has at most one solution $\lambda=\lambda(\varepsilon)$ when $|\varepsilon|$ and $\lambda$ are small enough. Thus, it is enough to show that $\frac{d}{d\lambda}F\neq 0$ when $|\varepsilon|$ and $\lambda\geq 0$ are small. The latter property of $F$ follows immediately from (\ref{gam}).
\qed

There exists a wide class of operators when Theorem \ref{nt1} and its proof can be simplified essentially. This can be done when there is a point $x_0\in X$ of a ``positive capacity" (or accessible) for the process $x(t)$, i.e.,
$$
\tau_{x_0}=\min(t:x(t)=x_0)<\infty
$$
for almost every initial point $x\in X$. In this case one can chose a killing potential with the support at the point $x_0$. For example, if $X$ is a countable set (i.e., $H_0$ is a lattice operator), one could take $q(x)=q_0\delta_{x_0}(x)$ where $q_0> 0$ and $\delta_{x_0}(x)=1$ when $x=x_0$, $\delta_{x_0}(x)=0$ when $x\neq x_0$. The best result will be obtained if $q_0=\infty$, i.e., $H_1$ is the operator $H_0$ defined on functions $\psi\in L^{2}(X,\mathcal{B,}\mu )$ with the Dirichlet boundary condition at the point $x_0$.

Note that the operator $H_1$ is well defined, and solutions of the corresponding Dirichlet problem with the boundary condition $\psi(x_0)=0$ can be obtained, for example, by the Kac-Feynman formula. The proof of Theorem \ref{nt1} remains valid with the relation $n_0=1$ following immediately from the fact that $H_1$ is the rank one perturbation of $H_0$. Thus the following theorem is valid.

\begin{theorem}\label{nt2}
Let conditions (a) and (b) hold, and let a point $x_0$ be accessible  for the process with the generator $H_0$. Let $H_1$ be the operator $H_0$ with the Dirichlet boundary condition $\psi(x_0)=0$.

Then the following estimate is valid for the number of non-positive eigenvalues of the operator $H_0-V$:
\begin{equation}\label{gt1}
N_0(V)\leq 1+\frac{1}{c(\sigma )}\int_{X}V(x)\int_{\frac{\sigma }{V(x)}%
}^{\infty }p_1(t,x,x)dt\mu (dx),~V\geq 0.
\end{equation}
\end{theorem}
We will illustrate the use Theorem \ref{nt2} with two simple examples.

\textbf{Example 1. One-dimensional Schr\"{o}dinger operator.} Let
$$
H=-\frac{d^2}{dx^2}-V(x) \quad \text {in}~~ L^2(R).
$$
Then the origin $x=0$ (and any other point) is accessible for the process with the generator $H_0$, and the estimate (\ref{gt1}) is valid with $p_1$ being the solution of the problem
$$
\frac{d}{dt}p_{1}=\frac{d^2}{dx^2}p_1, ~~~t>0;~~~p_1(t,0,y)=0,~~~p_1(0,x,y)=\delta_y(x).
$$
Then
$$
p_1(t,x,y)=\frac{e^{-\frac{(x-y)^2}{4t}}}{\sqrt{4\pi t}}-\frac{e^{-\frac{(x+y)^2}{4t}}}{\sqrt{4\pi t}},~~x,y>0.
$$
Thus
\begin{equation} \label{p11a}
p_1(t,x,x)=\frac{1-e^{-\frac{x^2}{t}}}{\sqrt{4\pi t}}.
\end{equation}
Similarly, the kernel $R^{(1)}_\lambda (x,y)$ of the resolvent $(-H_1-\lambda)^{-1}$ satisfies
$$
R^{(1)}_\lambda (x,y)=\frac{e^{-\sqrt\lambda|x+y|}-e^{-\sqrt\lambda|x-y|}}{2\sqrt\lambda}, \quad R^{(1)}_{+0} (x,x)=-|x|.
$$

Since $\int_0^\infty p_1(t,x,x)dt=-R^{(1)}_{+0}=|x|$, the estimate (\ref{gt1}) with $\sigma =0$ implies the Bargmann estimate (\ref{barg}). One also can obtain an estimate (we call it the refined Bargmann estimate) applicable to a slower decaying potentials by using (\ref{gt1}) with $\sigma >0$. Indeed, the formula for $p_1$ above implies that
$$
\int_{\frac{\sigma}{V(x)}}^\infty p_1(t,x,x)dt=|x|F(\frac{\sigma}{V(x)x^2}), \quad F(\gamma)=\int_\gamma^\infty \frac{1-e^{\frac{-1}{\tau}}}{\sqrt{4\pi\tau}}d\tau,
$$
and $F(\gamma)\leq 1$ for all $\gamma\geq 0;$ $F(\gamma)\leq \int_\gamma^\infty \frac{{1}}{\tau\sqrt{4\pi\tau}}d\tau=\frac{1}{\sqrt{\pi\gamma}} $ when $\gamma\geq 1.$ Thus,
\begin{equation} \label{rebarg}
N_0(V)\leq 1+\frac{1}{c(\sigma )}[\frac{1}{\sqrt{\sigma\pi}}\int_{x^2V(x)\leq\sigma} x^2 V^{3/2}(x)dx+\int_{x^2V(x)>\sigma} |x| V(x)dx],\quad d=1,
\end{equation}
with the same $c(\sigma)$ as in (\ref{1xx}). Note that the Bargmann estimate (\ref{barg}) does not provide any information on $N_0(V)$ in the case of the potential
\[
V(x)=O(\frac{1}{x^2 \ln|x|}), \quad |x|\to\infty,\quad d=1,
\]
(the integral in (\ref{barg}) diverges), while the refined formula (\ref{rebarg}) shows that $N_0(V)<\infty$ for this type of potentials.

\textbf{Example 2. One-dimensional lattice Schr\"{o}dinger operator.} Let
$$
H\psi(x)=-\Delta \psi-V(x) \psi=2\psi(x)-\psi(x+1)-\psi(x-1)-V(x)\psi(x) \quad \text {in}~~ L^2(Z).
$$
 The general solution of the equation $\Delta \psi-\lambda \psi =0,~\lambda >0,$ on the lattice $Z$ has the form $\psi =C_1a_1^x+C_2a_2^x$, where $a_{1,2}$ are the roots of the equation $a^2-(2+\lambda )a +1=0.$ If $a=\frac{2+\lambda+\sqrt{\lambda ^2+4\lambda}}{2},~\lambda >0,$ is the largest root, then the solution of the equation
$$
(\Delta -\lambda ) R_{\lambda}^{(0)}(x,y) =\delta (x-y)
$$
must have the form
$R_{\lambda}^{(0)}(x,y)=ca^{-|x-y|}$, where the constant $c$ can be easily found from the equation.  This leads to
\[
R_{\lambda}^{(0)}(x,y) =\frac{a^{1-|x-y|}}{2-(2+\lambda)a},\quad \lambda >0.
\]
If $H_1$ is the lattice Laplacian with the Dirichlet boundary condition at $x=0$ and
\begin{equation}\label{1212}
R_{\lambda}^{(1)}(x,y)=-\int_0^\infty p_1(t,x,y)e^{-\lambda t}dt
\end{equation}
is the kernel of its resolvent, then $R_{\lambda}^{(1)}(x,y)=R_{\lambda}^{(0)}(x,y)-R_{\lambda}^{(0)}(x,-y)$, and
\[
R_{\lambda}^{(1)}(x,x)=\frac{a-a^{1-2|x|}}{2-(2+\lambda)a},\quad \lambda >0.
\]
We note that $a \sim 1+ \sqrt{\lambda}$ and $2-(2+\lambda)a \sim -2\sqrt{\lambda}$ as $\lambda \to +0$. Hence,
$-R_{0}^{(1)}(x,x)=|x|$, and  therefore (\ref{1212}) and (\ref{gt1}) with $\sigma=1$ imply the Bargmann estimate for the one-dimensional lattice operator:
$$
N_0(V)\leq 1+\sum_Z |x| V(x).
$$

In order to obtain a refined Bargmann estimate in the lattice case, we note that
$$
p_1(t,x,y)=p_0(t,x,y)-p_0(t,x,-y), \quad \text{where}~~~p_0(t,x,y)=\frac{1}{2\pi}\int_{-\pi}^\pi e^{-2t(1-\cos\phi)+i(x-y)\phi}d\phi,
$$
i.e.,
$$
p_1(t,x,x)=p_0(t,x,x)-p_0(t,x,-x)=p_0(t,0,0)-p_0(t,2x,0).
$$
The integral above can be expressed through the modified Bessel function. This allows one to obtain the asymptotic behavior of $p_0(t,x,0)$ as $t,|x|\to\infty$. Another option is to apply Cramer's form of the central limiting theorem \cite{F} (Ch. 16, 7) which leads to the following result: if $t\to\infty$ then
$$
p_0(t,x,0)= \frac{ e^{-\frac{x^2}{4t}+O(\frac{|x|^4}{t^3})}}{\sqrt{4\pi t}}(1+O(\frac{1}{t})), \quad \text{for} \quad |x|\leq t^{2/3},
$$
$$
|p_0|\leq e^{-ct^{1/3}},~~~|x|\geq t^{2/3}.
$$
These formulas allow us to obtain the same estimate for $\int_\gamma^\infty p_1(t,x,x)dx$ as in the continuous case, which leads to
$$
N_0(V)\leq 1+ C_1(\sigma)\sum_{x:V(x)\leq\frac{\sigma}{x^2}}x^2V^{\frac{3}{2}}(x)+C_2(\sigma)\sum_{x:V(x)>\frac{\sigma}{x^2}}|x|V(x).
$$

\section{Operators on Riemannian manifolds}
This section is based significantly on the fundamental estimates \cite{grig}, \cite{grig-salov} for the heat kernel of the parabolic problems on Riemannian manifolds. The results of \cite{grig}, \cite{grig-salov} suit our goal perfectly in this section. We were not familiar with these estimates and got them directly for 2-D Laplacian and Bessel operators in a draft version of the paper (see \cite{arxiv}). The direct analytic approach may be useful since it provides an option to estimate (or to find explicitly) constants in all the formulas for $N_0(V)$, while the proofs in \cite{grig}, \cite{grig-salov} are non-constructive. Below we will illustrate this point when the lattice operator on $Z^2$ will be considered. The geometric methods in the spirit of \cite{grig}, \cite{grig-salov} are not applicable in this case.

The following symmetric operator in $L^2(R^d,\mu),~\mu(dx)=(1+|x|^2)^{\alpha/2}dx,$ will be considered in this section
\begin{equation}\label{lal}
L_\alpha=(1+|x|^2)^{-\alpha/2} \sum_{i=1}^d\frac{\partial}{\partial x_i}(1+|x|^2)^{\alpha/2}\frac{\partial}{\partial x_i}=
\Delta+\frac{\alpha x\cdot\bigtriangledown}{1+|x|^2},\quad d+\alpha>0.
\end{equation}
Its radial part
\begin{equation}\label{bal}
B_\alpha=\frac{\partial^2}{\partial r^2}+(\frac{d-1}{r}+\frac{\alpha r}{1+r^2})\frac{\partial}{\partial r}
\end{equation}
is close to the Bessel operator
$$
\widetilde{B}_{d+\alpha}=\frac{\partial^2}{\partial r^2}+\frac{d+\alpha-1}{r}\frac{\partial}{\partial r}
$$
when $r$ is large.

In contrast to the Bessel operator $\widetilde{B}_{d+\alpha}$, operator $L_\alpha$ does not have a singularity at $r=0$. Operator $L_\alpha$ represents the Laplacian on a smooth $d$-dimensional Riemannian manifold. More general (not spherically symmetrical) operators of the form
\begin{equation}\label{mgen}
L=\frac{1}{a}\rm{div} (a\nabla)=\Delta +\frac{\nabla a}{a}\cdot\nabla,~~
\quad c^-(1+|x|^\alpha)\leq a(x)\leq c^+(1+|x|^\alpha),
\end{equation}
are considered in  \cite{grig}, \cite{grig-salov}. For the sake of transparency, we will discuss here only the operator $L_\alpha$ given by (\ref{lal}). However, all the results of this section can be carried over to these more general operators (see the remark following the proof of Theorem \ref{trecc}).

The following estimate is proved in \cite{grig}, \cite{grig-salov} for the heat kernel $p_0(t,x,y)$ of the operator $L_\alpha$ ($p_0$ is the transition density with respect to the Lebesque measure $dx$ for the Markov process $x(t)$ with the generator  $L_\alpha$).
$$
p_0(t,x,y)\asymp\frac{e^{-c|x-y|^2/t}}{t^{d/2}(\sqrt t+1+|x|)^{\alpha/2}(\sqrt t+1+|y|)^{\alpha/2}}, \quad t> 0,~~x\in R^d.
$$
Here and below $f\asymp g$ means that $C_1g\leq f\leq C_2 g$ for some $C_1,C_2>0$. In particular,
\begin{equation}\label{p01}
p_0(t,x,x)\asymp \frac{1}{t^{d/2}(\sqrt t+1+|x|)^{\alpha}}, \quad t>0,~~x\in R^d.
\end{equation}
Hence, the process $x(t)$ is transient when $d+\alpha>2$, and it is recurrent when $d+\alpha\leq2$.

We are going to apply our general Theorem \ref{nt1} to the counting function $N_0(V)$ of the operator $H_\alpha=-L_\alpha-V,~V\geq 0$. Operators $L_\alpha$ and $H_\alpha$ are self-adjoint in the space $L^2(R^d,\mu)$ with the weight measure $\mu(dx)=(1+|x|^2)^{\alpha}dx$. The transition density $\widetilde{p}_0$ of the Markov process $x(t)$ in $R^d$ with respect to the Riemannian measure $\mu$ equals
\begin{equation}\label{p0w}
\widetilde{p}_0(t,x,y)=\frac{p_0(t,x,y)dy}{\mu(dy)}=\frac{p_0(t,x,y)}{(1+|y|^2)^{\alpha}}.
\end{equation}

The following proposition concerns the transient case ($d+\alpha >2$) and is a direct consequence of the standard CLR estimate (\ref{1xx}) (with $\sigma=1$, for simplicity) and estimate (\ref{p01}) for $p_0$. Denote $\langle x\rangle=2+|x|$ (the term $2$ on the right is chosen in order to allow the division by $\ln\langle x\rangle$).
\begin{theorem}\label{ttranz}
If $d+\alpha >2$, then the following estimates hold for $N_0(V)$.

If $d\geq 3$, then
$$
N_0(V)\leq C\left(\int_{\langle x\rangle^{2}V\leq 1}V^{(d+\alpha)/2}dx+\int_{\langle x\rangle^{2}V> 1}\langle x\rangle^{-\alpha}V^{d/2}dx\right).
$$

If $d=2$, then
$$
 N_0(V)\leq C\left(\int_{ \langle x\rangle^{2}V\leq 1}V^{(2+\alpha)/2}dx+\int_{\langle x\rangle^{2}V>1}\langle x\rangle^{-\alpha}V\ln(2\langle x\rangle^{2}V)dx\right).
$$

If $d=1$, then
$$
N_0(V)\leq C\left(\int_{ \langle x\rangle^{2}V\leq 1}V^{(1+\alpha)/2}dx+\int_{\langle x\rangle^{2}V>1}\langle x\rangle^{1-\alpha}Vdx\right).
$$

Here $C=C(d,\alpha)$ and $C\to\infty$ as $d+\alpha\to 2$.
\end{theorem}
\textbf{Proof.} Formula (\ref{1xx}) requires the operator $H$ to be self-adjoint. Thus, when it is applied to  $H_\alpha$, we need to use $\widetilde{p}_0$ in the formula, not $p_0$, and use the measure $\mu(dx)=(1+|x|^2)^{\alpha}dx$. From (\ref{p0w}) it follows that $\widetilde{p}_0(t,x,x)\mu(dx)=p_0(t,x,x)dx$. Thus using (\ref{1xx}) with $\sigma=1$ and (\ref{p01}), and making the substitution $\sqrt t=\langle x\rangle\tau$, we obtain that
\begin{equation*}
N_0(V) \leq C\int_{R^d}V(x)\int_{\frac{1 }{V(x)}%
}^{\infty }\frac{1}{t^{d/2}(\sqrt t+\langle x\rangle)^{\alpha}}dtdx=C\int_{R^d}\langle x\rangle^{2-d-\alpha}V(x)F(\frac{1}{\langle x\rangle\sqrt V})dx,
\end{equation*}
where
$$
F(\gamma)=\int_{\gamma}^\infty \frac{d\tau}{\tau^{d-1}(\tau+1)^\alpha}.
$$
It remains only to estimate $F$ separately for $\gamma \geq 1 $ and $\gamma <1$.
\qed

The next theorem concerns the recurrent case and is based on annihilation (Theorem \ref{nt1}). Thus we will assume that $d+\alpha\leq 2$ (the CLR estimate is meaningless in this case). We will exclude one-dimensional operators which can be studied by rank one perturbation technique (see Theorem \ref{nt2}). All the arguments below can be easily carried over to the case $d+\alpha> 2$  which would provide a better result than in the theorem above when $d+\alpha= 2+\varepsilon$ with small $\varepsilon>0$ (the constant $C$ will be bounded as $\varepsilon\to 0$).
\begin{theorem}\label{trecc}
If $d+\alpha \leq2$, then the following estimates hold for $N_0(V)$.

If $d+\alpha<2$ and $d\geq 3$, then
$$
N_0(V)\leq 1+ C\left(\int_{\langle x\rangle^{2}V\leq 1}\langle x\rangle^{4-2d-2\alpha}V^{2-\frac{d+\alpha}{2}}dx+\int_{\langle x\rangle^{2}V> 1}\langle x\rangle^{-\alpha}V^{d/2}dx\right).
$$

If $d+\alpha<2$ and $d=2$, then
$$
N_0(V)\leq 1+C\left(\int_{\langle x\rangle^{2}V\leq 1}\langle x\rangle^{-2\alpha}V^{2-\frac{d+\alpha}{2}}dx+\int_{\langle x\rangle^{2}V>1}\langle x\rangle^{-\alpha}V\ln(2\langle x\rangle^{2}V)dx\right).
$$

If $d+\alpha=2$ and $d\geq 3$, then
$$
N_0(V)\leq 1+ C\left(\int_{\langle x\rangle^{2}V\leq 1}V\frac{\ln^2\langle x\rangle}{\ln\frac{1}{V}}dx+\int_{\langle x\rangle^{2}V>1}[\langle x\rangle^{-\alpha}V^{d/2}+V\ln\langle x\rangle ]dx\right).
$$

If $d+\alpha=2$ and $d= 2$ (i.e., $\alpha=0$), then
\begin{equation}\label{rebarg2}
N_0(V)\leq 1+ C\left(\int_{\langle x\rangle^{2}V\leq 1}V\frac{\ln^2\langle x\rangle}{\ln\frac{1}{V}}dx+\int_{\langle x\rangle^{2}V>1}V\ln(\langle x\rangle^{3}V)dx\right).
\end{equation}
\end{theorem}
\textbf{Remark.} The estimates above can be simplified by making them a little rougher. In particular, the last estimate implies that
\begin{equation}\label{rn2}
N_0(V)\leq 1+ C\left(\int_{R^2}V(x)\ln\langle x\rangle dx+\int_{V>1}V(x)\ln V(x)dx\right), \quad d=2,~\alpha=0.
\end{equation}
\textbf{Proof.} We introduce a compactly supported spherically symmetrical killing potential $q=q(|x|),~q(|x|)>0$ for $|x|<R_0,~q(|x|)=0$ when $|x|\geq R_0$. In order to apply Theorem \ref{nt1}, we need an estimate on the fundamental solution $p_1(t,x,y)$ of the problem
$$
\frac{\partial p_1}{\partial t}=L_\alpha p_1-q(|x|)p_1, \quad p_1(0,x,y)=\delta_y(x).
$$

The following fundamental fact can be found in \cite{grig} (Theorem 10.10, parabolic case). We need the  Riemannian metric in which operator $L_\alpha$ is self-adjoint, see (\ref{lal}). Let $V(x,\sqrt t)$ be the Riemannian volume of the ball of radius $\sqrt t$ centered at $x$, and let $h=h(|x|)$ be the positive solution of the equation $L_\alpha h-qh=0$ ($h$ depends on $|x|$ in our case since the equation is spherically symmetric). Then
\begin{equation}\label{prec}
p_1(t,x,x)\asymp \frac {h^2(|x|)}{V(x,\sqrt t)h^2(\sqrt t+|x|)}, \quad t>0, ~x\in R^d.
\end{equation}

In our case (see \cite{grig-salov}, Theorem 4.9),
$$
V(x,\sqrt t)\asymp t^{d/2}(1+\sqrt t+|x|)^\alpha, \quad t>0,~x\in R^d.
$$
When $r\geq R_0$, function $h=h(r)$ satisfies the equation
$$
h''+(\frac{d-1}{r}+\frac{\alpha r}{1+r^2})h'=0
$$
from which it follows that
$$
h(r)=C_1\int_{R_0}^r\frac{du}{u^{d-1}(1+u^2)^{\alpha/2}}+C_2, \quad  r\geq R_0.
$$
Hence, for $r\to\infty$, we have
$$
h(r)\sim Cr^{2-d-\alpha} \quad \text{if} ~d+\alpha<2;\quad h(r)\sim C\ln r \quad \text{if} ~d+\alpha=2.
$$
Thus (\ref{prec}) implies the following two results
$$
p_1(t,x,x)\asymp \frac{\langle x\rangle^{2(2-d-\alpha)}}{t^{d/2}(\sqrt t+\langle x\rangle)^\alpha(\sqrt t+\langle x\rangle)^{2(2-d-\alpha)}},~~t>0,~x\in R^d, \quad d+\alpha<2,
$$
and
\begin{equation}\label{abcp1}
p_1(t,x,x)\asymp \frac{\ln^2\langle x\rangle}{t^{d/2}(\sqrt t+\langle x\rangle)^\alpha\ln^2(\sqrt t+\langle x\rangle)},~~t>0,~x\in R^d, \quad d+\alpha=2.
\end{equation}

Hence formula (\ref{gt}) with $\sigma=1$, after the substitution $\sqrt t=\tau \langle x\rangle$, implies that
\begin{equation}\label{abc}
N_0(V)\leq n_0+C\int_{R^d}\langle x\rangle^{2-d-\alpha}V(x)Fdx,
\end{equation}
where
$$
F=\int_{\gamma}^\infty \frac{d\tau}{\tau^{d-1}(\tau+1)^{4-2d-\alpha}},~~d+\alpha<2,
$$
$$F=\int_{\gamma}^\infty \frac{\ln^2\langle x\rangle d\tau}{\tau^{d-1}(\tau+1)^{\alpha}\ln^2([(\tau+1)\langle x\rangle]},~~d+\alpha=2.
$$
Here $\gamma=\frac{1}{\langle x\rangle\sqrt V}$. It will be shown below that $n_0=1$ if $q(x)$ is small enough. After that, the statements of the theorem follow from elementary estimates on  function $F$.

Let $d+\alpha<2$ and $d>2$. Obviously, $F<C\gamma^{d+\alpha-2}$ if $\gamma\geq1$, and $F<C\gamma^{2-d}$ if $\gamma<1$. This and (\ref{abc}) imply the first statement of the theorem (after the choice $n_0=1$ is justified). Let $d+\alpha<2$ and $d=2$, i.e., $\alpha<0$.
Then $F<C\gamma^{d+\alpha-2}$ if $\gamma\geq1$, and $F<C\ln(2/\gamma)$ if $\gamma<1$. This leads to the second statement of the theorem.

Let $d+\alpha=2.$ Then function $F$ for $\gamma\geq1 $ can be estimated as follows.
$$
F<C\int_{\gamma}^\infty\frac{\ln^2\langle x\rangle d\tau}{\tau\ln^2(\tau\langle x\rangle)}=C\int_{\frac{1}{\sqrt V}}^\infty\frac{\ln^2\langle x\rangle d\tau}{\tau\ln^2\tau}=2C\frac{\ln^2\langle x\rangle}{\ln\frac{1}{V}}.
$$
To estimate $F$ for $\gamma<1$, we split the interval of integration in the definition of $F$ into two parts, over intervals $(1,\infty)$ and $(\gamma,1)$. The integral over the first interval does not exceed $\ln\langle x\rangle$. The second one can be estimated by $C\gamma^{2-d}$ if $d>2$, or by $\ln(1/\gamma)<\ln(1/\gamma^2)$ if $d=2$. In particular, $F<\ln\langle x\rangle+\ln(1/\gamma^2)=\ln(\langle x\rangle^3V)$ if $\gamma<1,~d=2$. These estimates of $F$ imply the last two statements of the theorem.

In order to complete the proof of the theorem, it remains to show that $n_0=1$ if $q$ is small enough. To justify the choice of $n_0$, we fix a ball $B$ centered at the origin and containing the support of $q$, and impose the Neumann boundary condition at its boundary $\partial B$. Obviously, it is enough to show that the new problem (with zero Neumann data on $\partial B$) has at most one non-positive eigenvalue. The latter problem is the direct sum of the exterior and interior Neumann problems. The exterior problem is non-negative and can not have negative eigenvalues. It also does not have zero eigenvalue. Indeed, if $\psi\in L^2(R^d,\mu),~\mu(dx)=(1+|x|^2)^{\alpha/2}dx,$ is an eigenfunction of the exterior problem with zero eigenvalue, then
$$
0=\langle L_\alpha\psi,\psi\rangle_{L^2(R^d,\mu)}=\langle \nabla\psi,\nabla\psi\rangle_{L^2(R^d,\mu)},
$$
and therefore $\psi=0$. The interior problem with $q\equiv 0$ has only zero and positive eigenvalues. Since they depend continuously on $q$, one may have at most one non-positive eigenvalue when $q$ is small enough. Thus $n_0=1$ if $q$ is small enough.
\qed

\textbf{Remark.} The spherical symmetry of the operator $L_\alpha$ was used only to find explicitly the positive solution $h$ of the equation $L_\alpha h-qh=0$. The existence of this solution with appropriate estimates at infinity is proved in \cite{grig}, \cite{grig-salov} for non-symmetric operators (\ref{mgen}), and this allows one to extend Theorems \ref{ttranz}, \ref{trecc} to operators (\ref{mgen}).
\section{On the 2-D Schr\"{o}dinger operator.}
This section contains some comments and results concerning inequality (\ref{rn2}) in the classical (but still not well-understood) case of the 2-D Schr\"{o}dinger operator $H=-\Delta- V(x),~ V\geq 0,~ x\in R^2$.

First, let us note that the
constant $C$ in (\ref{rn2}) can not be specified since our proof of Theorem \ref{trecc} is based on the non-constructive estimates \cite{grig}, \cite{grig-salov} for the fundamental solution of the perturbed heat equation. In principle, one could find the constant in (\ref{rn2}) using a direct approach  suggested in \cite{arxiv} to study the heat equation. We will illustrate this direct approach below when the lattice operator on $Z^2$ is considered.

Let us discuss the relationship between (\ref{rn2}) and some other known results and conjectures.

Recently it was shown \cite{GNY} that the condition $N_0(V)<\infty$ implies that $V\in L^1(R^2)$, and moreover, there is a constant $c_0$ such that
\begin{equation}\label{gny}
N_0(V)\geq c_0\int_{R^1}Vdx.
\end{equation}
The proof of the latter estimate is rather difficult, while the implication
\begin{equation}\label{impl}
N_0(V)<\infty~\Rightarrow~V\in L^1(R^2)
\end{equation}
 can be justified more or less easily. We will show this in section 8 in the discrete case, but the continuous case can be treated similarly. We chose to consider the discrete case since the estimate (\ref{gny}) is not known for lattice operators, and one needs at least to justify (\ref{impl}). The statement converse to  (\ref{impl}) is not correct. Many counterexamples can be found in \cite{birL}, \cite{many}. Thus, our estimate (\ref{rn2}) is logarithmically close to an exact result.

One would like to have an estimate for $N_0(V)$ which is in agrement with quizi-classical asymptotics, i.e., an estimate such that $N_0(\alpha V)=O(\alpha)$ as $\alpha\to\infty$. Such an estimate can not be valid for general operators considered in the previous section, but it has been proved \cite{solO} for the 2-D Laplacian. The estimate is rather complicated and expressed in terms of local Orlich norms of the potential (an
earlier exposition of the author's technique and its development can be found in \cite{birS1}, \cite{birS2}).
Estimate (\ref{rn2}) is simpler, but logarithmically weaker when $\alpha\to \infty$. It implies $N_0(\alpha V)=O(\alpha\ln\alpha),~\alpha\to\infty$ (the estimate in the discrete case has the right scaling $O(\alpha)$, see section 6). Another simple estimate of  $N_0(V)$ can be found in \cite{st}. It is based on the Birman-Schwinger principle and is such that $N_0(\alpha V)=O(\alpha^2),~\alpha\to\infty$.

Estimate (\ref{rn2}) can be considered as a step in the direction of justification of the elegant conjecture \cite{many} which states that, in the case of the $2$-D Schr\"{o}dinger operator,
\begin{equation}\label{conm}
N_0( V)\leq 1+C_1\int_{r<r_0}V^\ast(r)\ln \frac{r_0}{r}dx+C_2\int_{r>r_0}V(x)\ln\frac{r}{r_0} dx+C_3\int_{R^2}V(x)dx,\quad r=|x|,
\end{equation}
where $r_0>0$ is arbitrary, and $V^\ast$ is the monotone spherical rearrangement of $V$, i.e.,  $V^\ast$ is the monotonic function such that the measures of the sets $\{x\in R^2:~V(x)\geq a>0\}$ and $\{x\in R^2:~V(|x|)\geq a>0\}$ coincide. Note that the validity of (\ref{conm}) for some $r_0$ implies its validity for each $r_0>0$ since from the rescaling $x\to x/r_0$ it follows that $N(\cdot)$ for the potentials $V(x)$ and $r_0^2V(x/r_0)$ coincide. Let us also mention that there exist numerous results concerning central potentials $V=V(r)$ \cite{seto}, \cite{newton}, \cite{many}, \cite{ls}. In particular, it is proved \cite{many} that if $V=V(r)$, then
\begin{equation}\label{conm11}
N_0( V)\leq 1+\int_{R^2}V(x)|\ln\frac{r}{r_0}| dx.
\end{equation}

Our result (\ref{rn2}) leads to

\begin{theorem}\label{td22}
Let $V(x)=0$ for $|x|>r_0$. Then
$$
N_0(V)\leq  1
+C\int_{|x|<r_0}V^\ast(r)\ln\frac{2r_0}{ r}dx.
$$
\end{theorem}
The proof is based on the following two lemmas.
\begin{lemma}\label{l52}
Inequality  (\ref{rn2}) implies that, for arbitrary V,
$$
N_0(V)\leq 1+C_1\int_{R^2}V(x)\ln\langle x\rangle dx+C_2\int_{V^\ast(r)>1}V^\ast(r)\ln\frac{\|V^\ast\|_1}{\pi r}dx.
$$
\end{lemma}
\textbf{Proof}. We note that
\begin{equation}\label{3211}
\int_{V>1}V(r)\ln Vdx=\int_{V^\ast>1}V^\ast(r)\ln V^\ast dx .
\end{equation}
Due to the Chebyshev inequality, for each $a\geq 1$,
$$
\pi (V^\ast)^2(a)=\text{meas} \{x:~V^\ast(x)>a\}\leq \frac{\|V^\ast\|_1}{a}.
$$
Thus,
\begin{equation}\label{3212}
\ln V^\ast(r)\leq\frac{1}{2}\ln\frac{\|V^\ast\|_1}{\pi r},
\end{equation}
which immediately leads to the statement of the lemma.
\qed
\begin{lemma}\label{l53}
Let $V(x)=0$ for $|x|>1$, and let $\overline{v}=\int_{|x|<1}Vdx\geq 1$. Then
$$
N_0(V)\leq  1+C\int_{|x|<1}V(x)dx
+C\int_{|x|<1}V^\ast(r)\ln\frac{1}{ r}dx.
$$
\end{lemma}
\textbf{Proof}. Consider the killing potential $q(x)=q_{\overline{v}}(x)=\overline{v}\chi(x),$ where $\chi(x)=1, |x|\leq 2;~\chi(x)=0, |x|> 2$. We apply formula (\ref{gt}) with $\sigma=1,~q=q_{\overline{v}}$ and $p_1(t,x,y)$ being the solution of (\ref{gta}) with $-H_1=\Delta-q_{\overline{v}}(x).$ Let us show that
\begin{equation}\label{intp1}
\int_{\frac{1}{V}}^\infty p_1(t,x,x)dt\leq C(1+\chi(x)\ln\frac{V}{\overline{v}}),\quad |x|\leq 1,
\end{equation}
with some $V$-independent constant $C$. Here $\chi(x)=1$ if $V(x)>\overline{v}$, and $\chi(x)=0$ otherwise.

We note that $p_1$ with $\overline{v}>1 $ does not exceed the value of the function $p_1$ with $\overline{v}=1.$ Thus, estimate (\ref{abcp1}) with $\alpha=0$ holds for $p_1$, and therefore
$$
\int_1^\infty p_1(t,x,x)dt\leq c_1,\quad |x|<1.
$$
Further, $p_1(t,x,y)\leq p_0(t,x,y)$, i.e.,
$$
p_1(t,x,y)\leq \frac {e^{-\frac{1}{4 t}}}{4\pi t}, \quad |x|=2,~|y|<1.
$$
We consider the restriction of $p_1$ to the disk $|x|\leq 2$. It satisfies 
the equation $u_t=\Delta u-q_{\overline{v}}u$ in the disk $|x|<2$, the initial data $u(0,x,y)=\delta(x-y),~|y|<1,$ and the boundary condition estimated above. This implies the following estimate:
$$
p_1(t,x,x)<\frac{e^{-t\overline{v}}}{4\pi t}+\frac {e^{-\frac{1}{4 t}}}{4\pi t}, \quad |x|\leq 1,
$$
which, for $\overline{v}>1$, leads to
$$
\int_{\frac{1}{V}}^1 p_1(t,x,x)dt<\int_{\frac{1}{V}}^\infty\frac{e^{-t\overline{v}}}{4\pi t}dt+\int_{\frac{1}{V}}^1
\frac {e^{-\frac{1}{4 t}}}{4\pi t}=\int_{\frac{\overline{v}}{V}}^\infty\frac{e^{-t}}{t}dt+\int_{\frac{1}{V}}^1
\frac {e^{-\frac{1}{4 t}}}{4\pi t}\leq c_2(1+\chi(x)\ln\frac{V}{\overline{v}}).
$$
This completes the proof of (\ref{intp1}).

Now, from (\ref{gt}), where the term $n_0$ can be estimated by the right-hand side in (\ref{conm11}), and (\ref{intp1}) it follows that
$$
N_0(V)\leq 1+C\int_{|x|<1}V(x)dx+C\int_{x:V(x)>\overline{v}}V(x)\ln  \frac{V}{\overline{v}}dx.
$$
In order to complete the proof, it remains to note that by (\ref{3211}), (\ref{3212}) one can rewrite the last integral above as follows
$$
\int_{x:V(x)>\overline{v}}V(x)\ln  \frac{V}{\overline{v}}dx=\int_{x:V^*(r)>\overline{v}}V^*(r)\ln  \frac{V^*(r)}{\overline{v}}dx\leq\frac{1}{2}\int_{|x|<1}V^*(r)\ln  \frac{1}{\pi r}dx.
$$
\qed

\textbf{Proof of Theorem \ref{td22}.} If $||V||\leq1$, then the statement of the theorem follows immediately from Lemma \ref{l52}. If  $||V||>1$, then the statement follows from Lemma \ref{l53} after the rescalling $x\to x/r_0, ~V(x)\to r_0^2V(x/r_0)$.\qed
\section{A two-dimensional lattice operator}

We consider two-dimensional lattice operators
\begin{equation} \label{bargZo}
H\psi(x)=-\Delta \psi-V(x) \psi \quad \text {in}~~ L^2(Z^2)
\end{equation}
in this section. The main results of the section are stated in the following two theorems. The first theorem justifies a Bargmann type estimate and a refined one for $N_0(V)$. Both estimates follow from Theorem \ref{nt2} and need an estimate on $p_1(t,x,x)$ for their proofs. The second theorem is a limit theorem for the random variable $\tau_x=\min\{t:x(t)=0\}$ which is the first time when the random walk on $Z^2$ starting at $x$ visits the origin.

\begin{theorem} \label{t1}
The following estimate holds for operator (\ref{bargZo}):
\begin{equation}\label{tran1z}
N_0(V)\leq  1+C\sum_{Z^2} \ln(2+|x|) V(x),
\end{equation}
which is a particular case ($\sigma=0$) of more general estimates which are valid for
all $\sigma\geq 0$
\begin{equation} \label{rebarg2l}
N_0(V)\leq 1+C_1(\sigma)\sum_{x:V(x)<\frac{\sigma}{\langle x\rangle}} \frac{V(x)}{\ln\frac{\sigma}{V(x)}}\ln^2\langle x\rangle
+C_2(\sigma)\sum_{x:V(x)\geq\frac{\sigma}{\langle x\rangle}}V\ln\langle x\rangle, \quad \langle x\rangle=2+|x|.
\end{equation}
\end{theorem}
\textbf{Remarks. 1.} Formulas (\ref{tran1z}), (\ref{rebarg2l}) are in agrement with the quasi-classical asymptotics, i.e., $N_0(\alpha V)=O(\alpha)$ as $\alpha\to\infty$. The main difference between (\ref{rebarg2l}) and (\ref{rebarg2}) is that the second integrand in
 (\ref{rebarg2}) contains $V$ under the logarithm sign which is absent in (\ref{rebarg2l}). Its presence in  (\ref{rebarg2}) is due to the non-integrability of the transition density $p_0(t,x,x)$ at $t=0$ for the Laplacian in $R^2$.

\textbf{ 2.}  After  (\ref{rebarg2l}) is proved, one can get a better estimate:
\begin{equation} \label{rebarg2la}
N_0(V)\leq 1+m+C_1(\sigma)\sum_{x:V(x)<\frac{\sigma}{\langle x\rangle}} \frac{V(x)}{\ln\frac{\sigma}{V(x)}}\ln^2\langle x\rangle
+C_2(\sigma)\sum_{x:h>V(x)\geq\frac{\sigma}{\langle x\rangle}}V\ln\langle x\rangle,
\end{equation}
where $m=\#\{x:V(x)\geq h\}$ and $h$ is an arbitrary constant (or function). Indeed, let us introduce the potential $\widetilde{V}(x)$ which coincides with $V$ at the points $x$ where $V(x)<h$, and $\widetilde{V}(x)=0$ elsewhere. The operators $H$ with the potentials $V$ and $\widetilde{V}$ differ by an operator of rank $m$, and the difference between the numbers of their eigenvalues can be at most $m$. Thus estimate (\ref{rebarg2l}) for the potential $\widetilde{V}$ implies (\ref{rebarg2la}).

In order to prove the theorem, we will need three lemmas.

Let $R_{\lambda}^{(0)}(x,y),$ $ R_{\lambda}^{(1)}(x,y)$ be the kernels of the resolvents of the operators $\Delta =-H_0$ and $-H_1$, respectively, where $-H_1$ is obtained from $-H_0$ by imposing the Dirichlet boundary condition at the origin (annihilation of the Markov process at this point). Obviously,
\[
R_{\lambda}^{(1)}(x,y)=-\int_0^{\infty}p_1(t,x,y)e^{-\lambda t}dt,~~\lambda >0,
\]
where $p_1$ is the transition probability for the Markov process with the generator $-H_1$. The inverse Laplace transform implies
\begin{equation} \label{invL}
p_1(t,x,y)=-\int_{\text{Re}\lambda=a}R_{\lambda}^{(1)}(x,y)e^{\lambda t}d\lambda \quad \text{for any} ~~a>0.
 \end{equation}
The next two lemmas concern the asymptotics of $R_{\lambda}^{(0)}$ and $ R_{\lambda}^{(1)}$ as $\lambda\to 0$ which are needed to prove the third lemma providing an estimate on $p_1(t,x,x)$.

\begin{lemma} \label{ldiscr}
The following asymptotic expansion of the resolvent of the operator $H_0=-\Delta$ in $L^2(Z^2)$ holds uniformly in $x\in Z^2$.
$$
R_{\lambda}^{(0)}(x,0)=\frac{1}{4\pi}\ln (\lambda(1+|x|)^2)+u(x)+O(\lambda(1+|x|)^2\ln\frac{1}{\lambda}),~ \quad \lambda\to+0,~~~ |u(x)|<C,
$$
where $\alpha=u(0)$ is real. (Here and below $F=O(f)$ means that $|F|\leq C|f|$)
\end{lemma}
\textbf{Proof.}
The Fourier method applied to the equation $(\Delta-\lambda)\psi=\delta(x)$ leads, for $\lambda>0$, to
$$
R_{\lambda}^{(0)}(x,0)=\frac{1}{(2\pi)^2}\int_{[\pi,\pi]^2}\frac{e^{i(x,\phi)}d\phi}{2\cos\phi_1+2\cos\phi_2-4-\lambda}
$$
\begin{equation} \label{resla1}
=
\frac{-1}{(2\pi)^2}\int_{[-\pi,\pi]^2}\frac{e^{i(x,\phi)}d\phi}{4\sin^2\frac{\phi_1}{2}+4\sin^2\frac{\phi_2}{2}+\lambda},
\end{equation}
where $\phi=(\phi_1,\phi_2)\in [-\pi,\pi]^2 \subset R^2.$ We put here
$$
4\sin^2\frac{\phi_1}{2}+4\sin^2\frac{\phi_2}{2}=|\phi|^2+h(\phi),~|h(\phi)|<C|\phi|^4.
$$
The difference between (\ref{resla1}) and the same integral with $h(\phi)=0$ is
\[
v(\lambda,x)=\frac{1}{(2\pi)^2}\int_{[-\pi,\pi]^2}\frac{h(\phi)e^{i(x,\phi)}d\phi}
{[4\sin^2\frac{\phi_1}{2}+4\sin^2\frac{\phi_2}{2}+\lambda][|\phi|^2+\lambda]}.
\]
The latter integral converges to a bounded function $v(0,x)$
as $\lambda\to +0$. Moreover, the difference $v(\lambda,x)-v(0,x)$ is  bounded by $C\lambda\ln\frac{1}{\lambda}$ as $\lambda \to +0$, uniformly in $x\in Z^2$. Hence, the following relation holds uniformly in $x\in Z^2$:
\[
R_{\lambda}^{(0)}(x,0)=\frac{-1}{(2\pi)^2}\int_{[-\pi,\pi]^2}\frac{e^{i(x,\phi)}d\phi}{|\phi|^2+\lambda}+
v(0,x)+O(\lambda\ln\frac{1}{\lambda}), \quad
\lambda\to +0,~~ \quad |v(0,x)|<C.
\]
This implies that, uniformly in $x\in Z^2$,
\begin{equation} \label{11aa}
R_{\lambda}^{(0)}(x,0)=\frac{-1}{(2\pi)^2}\int_{|\phi|<1}\frac{e^{i(x,\phi)}d\phi}{|\phi|^2+\lambda}+
w(x)+O(\lambda\ln\frac{1}{\lambda}), \quad
\lambda\to +0,~~  \quad |w(x)|<C.
\end{equation}

We represent the integral above as $I_1+I_2+I_3$, where
$$
I_1=\int_{|\phi|<1}\frac{d\phi}{|\phi|^2+\lambda},~~I_2=\int_{|\phi|<1}\frac{(e^{i(x,\phi)}-1)d\phi}{|\phi|^2},~~
I_3=\int_{|\phi|<1}\frac{\lambda(1-e^{i(x,\phi)})d\phi}{(|\phi|^2+\lambda)|\phi|^2}.
$$
Obviously, $I_1=\pi[\ln(1+\lambda)-\ln\lambda]$. In order to evaluate $I_2$, we note that it depends only on $r=|x|$. We replace $x$ in the formula for $I_2$ by $x=(r,0)$, differentiate with respect to $r$, and pass to the polar coordinates $\sigma=|\phi|,~ \theta=\arctan \phi_2/\phi_1$:
\[
\frac{dI_2}{dr}=i\int_{|\phi|<1}\frac{\phi_1e^{ir\phi_1}d\phi}{|\phi|^2}=i\int_0^{2\pi}\int_0^1 \cos \theta e^{ir\sigma\cos \theta }d\sigma d\theta=\frac{1 }{r}\int_0^{2\pi}(e^{ir\cos \theta }-1)d\theta.
\]
Thus, $\frac{dI_2}{dr}(r)=\frac{-2\pi}{r}+O(\frac{1}{r^{3/2}}),~r\to\infty$, and therefore $I_2=-2\pi\ln(1+r)+O(1)$. Finally,
$$
I_3\leq C\int_{|\phi|<1}\frac{\lambda|x|^2|\phi|^2d\phi}{(|\phi|^2+\lambda)|\phi|^2}=
C\int_{|\phi|<1}\frac{\lambda|x|^2d\phi}{(|\phi|^2+\lambda)}\leq C_1\lambda|x|^2\ln\frac{1}{\lambda},~~\lambda\to +0.
$$
These estimates for $I_j,~j=1,2,3,$ and (\ref{11aa}) justify the statement of the lemma (one can easily check that $\alpha$ is real).
\qed

\begin{lemma} \label{ldiscr1}
The following asymptotic expansion for the resolvent of the operator $H_1$ in $L^2(Z^2)$ holds uniformly in $x\in Z^2$ when $\lambda\to+0$ and $\lambda (1+|x|)^2\ln(2+|x|)\leq 1$:
$$
R_{\lambda}^{(1)}(x,x)=-\frac{1}{\pi}\ln (1+|x|)-2v(x)-\frac{[\frac{1}{2\pi}\ln (1+|x|)+v(x)]^2}{\frac{1}{4\pi}\ln\lambda+\alpha}+O(\lambda(1+|x|)^2\ln\frac{1}{\lambda}),
$$
where $v(x)=u(x)-\alpha,~~|v(x)|<C$ and $u(x),~\alpha$ are defined in the previous lemma.
\end{lemma}

 \textbf{Proof.}
Since $\Delta R_{\lambda}^1(x,y)=0,~\lambda >0,$ if $x\neq y$ and $x\neq 0$,
the kernel $R_{\lambda}^1(x,y),~\lambda >0,$ must have the form  $R_{\lambda}^1(x,y)=R_{\lambda}^0(x,y)+cR_{\lambda}^0(x,0)$, where $c=c(y)$ can be found from the condition $R_{\lambda}^1(0,y)=0$. This immediately implies
\begin{equation}\label{rl1}
R_{\lambda}^{(1)}(x,x)=R_{\lambda}^{(0)}(x,x)-\frac{[R_{\lambda}^{(0)}(x,0)]^2}{R_{\lambda}^{(0)}(0,0)}=
\frac{[R_{\lambda}^{(0)}(0,0)]^2-[R_{\lambda}^{(0)}(x,0)]^2}{R_{\lambda}^{(0)}(0,0)},~~\lambda>0.
\end{equation}
It remains only to put here the expansion for $R_\lambda^{(0)}$ from the previous lemma.
\qed

\begin{lemma} \label{ldiscrt}
The following estimate for the heat kernel $p_1$ holds uniformly in $t\geq 0$ and $x\in Z^2$:
\begin{equation} \label{p1d}
 p_1(t,x,x)\leq C\frac{\ln ^2(2+|x|)}{(1+t)\ln ^2(2+t+|x|)}.
\end{equation}

\end{lemma}
\textbf{Proof.} Since $p_1(t,x,x)\leq p_0(t,x,x)\leq \frac{c}{1+t}$, estimate (\ref{p1d}) holds when $t\leq (1+|x|)^4$. Let us prove it for $t>(1+|x|)^4$.

Lemma \ref{ldiscr1} allows us to rewrite (\ref{invL}) in the form
\begin{equation} \label{aGa2}
p_1=-\int_{\Gamma} R_{\lambda}^{(1)}(x,y)e^{\lambda t}d\lambda,
\end{equation}
where the contour $\Gamma$ consists of the bisectors of the third and second quadrants of the $\lambda-$plane with the  direction on $\Gamma$ such that Im$ \lambda$ increases  along $\Gamma$.

Obviously,
\begin{equation} \label{t1t}
\int_\Gamma e^{\lambda t}d\lambda=0 \quad \text{for}~~ t>0.
\end{equation}
Further, replacing $\Gamma$ by a contour $\gamma$ around the negative semi-axis in the $\lambda-$plane, we obtain that, for any real $\beta$,
\[
\int_{\Gamma}\frac{1}{\ln\lambda +\beta} e^{\lambda t}d\lambda=\int_{\gamma}\frac{1}{\ln\lambda+\beta} e^{\lambda t}d\lambda
=\int_0^{\infty}[\frac{1}{\ln\sigma+\pi i+\beta}-\frac{1}{\ln\sigma-\pi i+\beta}]e^{-\sigma t}d\sigma
\]
\begin{equation} \label{tt}
=\int_0^{\infty}\frac{2\pi i}{(\beta+\ln\sigma)^2+\pi^2}e^{-\sigma t}d\sigma\asymp\frac{1}{(1+t)\ln^2(2+t)}
,~~t>2.
\end{equation}
The last two relations show that the contributions to $p_1$ in (\ref{aGa2}) from the main terms of asymptotics of $R_{\lambda}^{(1)}$ (see Lemma \ref{ldiscr1}) satisfy (\ref{p1d}) when $t>1+|x|^4$. It remains to check that the remainder term $\rho_\lambda$ of the asymptotics also has this property.

To estimate the integral of the remainder term, we split the contour of integration into two parts  $\Gamma_1=\Gamma\bigcap \{\lambda:|\lambda|\leq \tau\}$ and $\Gamma_2=\Gamma\backslash \Gamma_1,$ where $\tau=c(1+|x|)^{-3},~c=\min\frac{1+|x|}{\ln(2+|x|)}$. Then the estimate of $\rho_\lambda$ provided by Lemma \ref{ldiscr1} is valid on $\Gamma_1$ and, for $\varepsilon\in(0,1/2)$, we have
$$
|\int_{\Gamma_1} \rho_{\lambda}e^{\lambda t}d\lambda|\leq C_\varepsilon|\int_{\Gamma_1} \lambda ^{1-\varepsilon}(1+|x|)^2e^{\lambda t}d\lambda|\leq \varepsilon|\int_{\Gamma} \lambda ^{1-\varepsilon}(1+|x|)^2e^{\lambda t}d\lambda|=C_{1,\varepsilon}\frac{(1+|x|)^2}{t^{2-\varepsilon}}.
$$
The right-hand side here can be estimated from above by the right-hand side in  (\ref{p1d}) when $t>(1+|x|)^4$. To get a similar estimate for the integral over $\Gamma_2$, we note that $R_{\lambda}^{(1)}(x,x)\leq C(1+|x|)^{3},~\lambda\in \Gamma_2,$ since the norm of the resolvent does not exceed the inverse distance from the spectrum. The same estimate of the resolvent could be obtained from (\ref{resla1}), (\ref{rl1}). Since the main terms of the asymptotics of $R_{\lambda}^{(1)}(x,x)$ do not exceed $C\ln^2(2+|x|)$, it follows that $\rho_{\lambda}\leq C(1+|x|)^{3},~\lambda\in \Gamma_2.$ Hence
$$
|\int_{\Gamma_2} \rho_{\lambda}e^{\lambda t}d\lambda|\leq C(1+|x|)^{3}|\int_{\Gamma_2}e^{\lambda t}d\lambda|\leq C\frac{(1+|x|)^{3}}{t}e^{\frac{-\sqrt 2ct}{2(1+|x|)^{3}}},
$$
and the latter expression can be estimated from above by the right hand side in  (\ref{p1d}) when $t>(1+|x|)^4$.
\qed

\textbf{ Proof of Theorem \ref{t1}.}  One simply needs to use  (\ref{gt1}) with estimate (\ref{p1d}) for $p_1$. Indeed, if $V\leq \frac{\sigma}{\langle x\rangle}$, then
\begin{equation} \label{98}
\int _{\frac{\sigma}{V}}^\infty p_1(t,x,x)dt\leq C\int _{\frac{\sigma}{V}}^\infty \frac{\ln^2\langle x\rangle dt}{(1+t)\ln^2(t+\langle x\rangle)}\leq C \int _{\frac{\sigma}{V}}^\infty \frac{\ln^2\langle x\rangle dt}{t\ln^2t}=C\frac{\ln^2\langle x\rangle }{\ln\frac{\sigma}{V}}.
\end{equation}
It remains to show that $\int _{\frac{\sigma}{V}}^\infty p_1(t,x,x)dt\leq C\ln\langle x\rangle$ when  $V\geq \frac{\sigma}{\langle x\rangle}$. To justify the latter inequality, we split the integral into two terms: over intervals $(\frac{\sigma}{V},\langle x\rangle)$ and $(\langle x\rangle, \infty)$. The second term can be estimated by referring to (\ref{98}) with $\frac{\sigma}{V}=\langle x\rangle$. The first term can be estimated by omitting $t$ under the logarithm sign in (\ref{p1d}).
\qed

The following limit theorem holds for the random variable $\ln \tau_x$, where $\tau_x$ is the first time when the random walk on $Z^2$ starting at $x$ visits the origin.
\begin{theorem}
The following relation holds for each $\alpha\in R$:
$$
P_x\{\frac{\ln\tau_x}{\ln|x|}\leq \alpha\}\rightarrow \frac{(\alpha-2)_+}{\alpha} \quad \text{as} \quad |x|\to \infty.
$$
Here $(\alpha-2)_+=\max(0,\alpha-2)$.
\end{theorem}
\textbf{Proof.} Let $q_x(t)$ be the density of the random variable $\tau_x$. It satisfies the relations
$$
\frac{\partial}{\partial t}q_x(t)=\Delta q_x(t),\quad t>0,~x\neq 0;\quad q_0(t)=1,~~t\geq 0;\quad q_x(0)=0,~~x\neq 0.
$$
Hence, the Laplace transform of $q_x(t)$ can be expressed through the resolvent $R_{\lambda}^{(0)}$:
\begin{equation} \label{abab}
E_xe^{-\lambda \tau}=\int_0^\infty e^{-\lambda t}q_x(t)dt=\frac{R_{\lambda}^{(0)}(x,0)}{R_{\lambda}^{(0)}(0,0)}.
\end{equation}
 The latter formula and Lemma \ref{ldiscr} after the rescaling $\lambda=\lambda_1|x|^{-\alpha}$, $\alpha>2,$ imply
$$
E_xe^{-\lambda_1\frac{ \tau}{|x|^\alpha}}\rightarrow \frac{\alpha-2}{\alpha} \quad \text{as} \quad |x|\to \infty.
$$
Since the right-hand side above does not depend on $\lambda_1$, the continuity theorem for the Laplace transform of probability measures on the positive half-line provides the following result. If $|x|\to\infty$, then the distribution of the random variable $\tau_x/|x|^\alpha$ converges weakly over space $C_\text{com}(R_+^1)$ to the measure which has a single atom of mass $\frac{(\alpha-2)_+}{\alpha}$ at the point $t=0$, i.e., for each $\varepsilon>0$ and $\alpha>2$,
$$
P_x\{\frac{\tau}{|x|^\alpha}\leq \varepsilon\}\to\frac{\alpha-2}{\alpha},~~P_x\{\frac{\tau}{|x|^\alpha}\geq \varepsilon^{-1}\}\to\frac{2}{\alpha},~~|x|\to\infty.
$$
These relations imply the statement of the theorem for $\alpha>2$, and therefore for each $\alpha\in R$.
\qed

\section{General discrete Schr\"{o}dinger operators with recurrent underlying Markov processes}

This section is devoted to a Bargmann type estimate for general lattice operators. We also will show here that shift-invariant estimates of the form (\ref{1xx}) can not be valid for operators with recurrent underlying Markov processes.

Let $X$ be a countable set and let $H_0$ be a symmetric non-negative operator on $L^2(X)$ with matrix elements $h(x,y)$, i.e.
\begin{equation}\label{above12}
H_0\psi(x)=\sum_{y\in X}h(x,y)\psi(y), \quad h(x,y)=h(y,x).
\end{equation}
It is assumed that
\begin{equation}\label{above1}
h(x,y)\leq 0~~\text{if}~~x\neq y,\quad \sum_{y\in X}h(x,y)= 0 ;\quad h(x,x)\leq c_0 ~~~\text {for all} ~~x\in X.
\end{equation}
Obviously, operator $H_0$ can be written in the form
$$
H_0\psi(x)=\sum_{y\in X:y\neq x}h(x,y)(\psi(y)-\psi(x)).
$$
The first two conditions in (\ref{above1}) guarantee the existence and uniqueness of the Markov process $x(t)$ with the generator $-H_0$ and that the operator $H_0$ is non-negative. The last condition in (\ref{above1}) is needed to avoid a pathological behavior of the Markov process $x(t)$. We also assume connectivity, i.e., $X$ can not be split in two disjoint non-empty sets $X_1\cup X_2$ in such a way that $h(x_1,x_2)=0$ for each $x_1\in X_1,~x_2\in X_2$. If $H_0=-\Delta$ on $Z^d$, we have $h(x,x)=2d,~h(x,y)=-1$ when $|x-y|=1, ~h(x,y)=0$ when $|x-y|>1$.

Let $p_0(t,x,y)$ be the transition probability, i.e., $p_0$ is the kernel of the Markov semigroup $e^{-tH_0}$, and let
\begin{equation}\label{rp}
R^{(0)}_\lambda(x,y)=-\int_0^\infty  p_0(t,x,y)e^{-\lambda t}dt
\end{equation}
be the kernel of the resolvent $R^{(0)}_\lambda=(-H_0-\lambda)^{-1}$ of the operator $-H_0$. The connectivity assumption implies that $p_0(t,x,y)>0$ and $R_\lambda<0$ for all the values of the arguments. Since $-H_0\leq 0,$ the function $R^{(0)}_\lambda(x,y)$ is analytic in $\lambda\notin (-\infty,0]$.

We assume that the process $x(t)$ is recurrent. Hence
\begin{equation}\label{recu}
\int_0^\infty p_0(t,x,x)dt=\infty,
\end{equation}
due to Proposition \ref{pr4}. The latter relation implies that
\begin{equation}\label{rinf}
\lim_{\lambda \to +0}R^{(0)}_\lambda(x,x)=-\infty.
\end{equation}

We fix a point $x_0\in X$. Let $p_1$ be the transition probability of the Markov process with the generator $-H_1$ obtained from $-H_0$ by imposing the Dirichlet boundary condition at the point $x_0$ (annihilation of the Markov process with the generator $-H_0$ at this point). Consider the kernel of the resolvent of the operator $-H_1$:
\begin{equation} \label{res1abd}
R_{\lambda}^{(1)}(x,y)=-\int_0^{\infty}p_1(t,x,y)e^{-\lambda t}dt,~~\lambda >0.
\end{equation}
Similarly to (\ref{rl1}), we have
\begin{equation} \label{res1ab}
R_{\lambda}^{(1)}(x,x)=R_{\lambda}^{(0)}(x,x)-\frac{[R_{\lambda}^{(0)}(x,x_0)]^2}{R_{\lambda}^{(0)}(x_0,x_0)},~~\lambda>0,
\end{equation}
and
\begin{equation} \label{res1}
\widetilde{R}(x,x_0):=-R_{0}^{(1)}(x,x)=\lim_{\lambda \to +0}[\frac{[R_{\lambda}^{(0)}(x,x_0)]^2}{R_{\lambda}^{(0)}(x_0,x_0)}-R_{\lambda}^{(0)}(x,x)]<\infty.
\end{equation}
 The finiteness of the expression above follows from  Proposition \ref{pr4} and the statement on transitivity in Theorem \ref{nt1}.
In some cases, expression (\ref{res1}) can be written in a simpler form:
\begin{equation} \label{res}
\widetilde{R}(x,x_0)=2\lim_{\lambda \to +0}[R_{\lambda}^{(0)}(x,x_0)-R_{\lambda}^{(0)}(x_0,x_0)].
\end{equation}
For example, the latter formula is valid when operator $H_0$ is translation-invariant with respect to some transitive group on $X$ (and the Markov process with the generator $-H_0$ is recurrent and connected). Indeed, (\ref{res1ab}) can be rewritten in the form
$$
R_{\lambda}^{(1)}(x,x)=\frac{B(R_{\lambda}^{(0)}(x_0,x_0)+R_{\lambda}^{(0)}(x,x_0))}
{R_{\lambda}^{(0)}(x_0,x_0)},
$$
where $B=R_{\lambda}^{(0)}(x_0,x_0)-R_{\lambda}^{(0)}(x,x_0)$. Since $R_{\lambda}^{(0)}(x_0,x_0)$ and $R_{\lambda}^{(0)}(x,x_0)$ have the same sign (they are negative), the ratio satisfies
$$\frac
{R_{\lambda}^{(0)}(x_0,x_0)}{R_{\lambda}^{(0)}(x_0,x_0)+R_{\lambda}^{(0)}(x,x_0)}\in[0,1].
$$
Hence $B$ is bounded when $\lambda\to +0$ (since $R_{0}^{(1)}(x,x)$ is bounded). From here and $\lim_{\lambda \to +0}|R_{\lambda}^{(0)}(x_0,x_0)|=\infty, ~|R_{0}^{(1)}(x,x)|<\infty$ it follows that
$$
R_{0}^{(1)}(x,x)=\lim_{\lambda \to +0}\frac{[R_{\lambda}^{(0)}(x_0,x_0)]^2-[R_{\lambda}^{(0)}(x,x_0)]^2}{R_{\lambda}^{(0)}(x_0,x_0)}=\lim_{\lambda \to +0}\{B[2-\frac{B}{R_{\lambda}^{(0)}(x_0,x_0)}]\}=2\lim_{\lambda\to +0}B,
$$
i.e., (\ref{res}) holds.

Due to (\ref{res1abd}) and (\ref{res1}), the following Bargmann type estimate for the lattice operator $H=H_0-V(x)$ follows directly from Theorem \ref{nt2} (with $\sigma=0$) and the arguments in Remark 2 following Theorem \ref{t1}.
\begin{theorem}\label{barggen}
Let the Markov process $x(t)$ with the generator $-H_0$, defined by (\ref{above12}),  be recurrent, and let the connectivity condition hold. Then

1) the function $\widetilde{R}$
is finite for all $x,x_0\in X$ and positive for $x\neq x_0$ (it vanishes if $x=x_0$),

2) the following estimate holds
\begin{equation}\label{f2}
N_0(V)\leq 1+ \sum_{x\in X}\min(1,V(x)\widetilde{R}(x,x_0)).
\end{equation}
\end{theorem}

Proposition \ref{tno} below shows that a space invariant estimate of $N_0(V)$ can not be valid for the discrete operator $H$ with a recurrent underlying Markov process, but first we need to prove the following preliminary result.
\begin{proposition}\label{tinter}
For each $y\in X$ and $\alpha>0,$ the operator $H=H_0-\alpha\delta_{y}(x)$ has a unique simple negative eigenvalue $\lambda=\lambda (y,\alpha)<0$.
\end{proposition}
\textbf{Proof.} The uniqueness is due to the fact that $H$ is a rank one perturbation of  $H_0$. Let us show the existence of the eigenvalue. First we note that $\sum_x p_0(t,x,y)=1$, and therefore (\ref{rp}) implies that
\begin{equation}\label{sum1}
 \sum_x R^{(0)}_\lambda(x,y)=\frac{-1}{\lambda}, \quad\lambda>0.
\end{equation}
Formula (\ref{rp}) implies also that $R^{(0)}_\lambda(x,y)< 0$. Thus from (\ref{sum1}) it follows that $|R^{(0)}_\lambda(x,y)|\leq \frac{1}{\lambda}$ for each $x,y\in X$ and $\lambda>0$. This and (\ref{sum1}) lead to the estimate $\sum_x [R^{(0)}_\lambda(x,y)]^2\leq \frac{1}{\lambda^2},$ i.e., $R^{(0)}_\lambda(x,y)\in L^2(X),~y\in X$.

We look for an eigenfunction in the form $\psi_\lambda(x)=R^{(0)}_\lambda(x,y),~\lambda>0$. Since $(-H_0-\lambda)\psi_\lambda=\delta_{y}(x)$, $\psi_\lambda$ will be an eigenfunction of $H=H_0-\alpha\delta_{y}(x)$ with the eigenvalue $-\lambda$ if $-\alpha R^{(0)}_\lambda(y,y)=1$. The latter equation has a solution $\lambda=\lambda (y,\alpha)>0$ for each $y\in X$ and $\alpha>0$ due to (\ref{rinf}) and the relation $\lim_{\lambda\to \infty} R^{(0)}_\lambda(y,y)=0.$
\qed

The next theorem shows that for each $\gamma$ one can find a potential $V\geq 0$ such that $\sum_{x\in X} V^\gamma (x)$ is arbitrarily small and the operator $H$ has infinitely many negative eigenvalues. Hence, estimate (\ref{c}) can not be valid for the operator $H$. The potential $V$ will be constructed when a uniformity condition on the unperturbed operator $H_0$ holds. We assume that there exists a metric $d(x,y)$ on $X$ (for example, $l^1-$metric on $Z^d$) such that the following two relations hold:

a) $|R^{(0)}_\lambda(x,x)|\geq \beta(\lambda)$ for $\lambda>0$ and some $\beta(\lambda)>0$, and $\beta(\lambda)\to \infty$ as $\lambda\to+0$;

b) $\sum_{x:d(x,y)>r}| R^{(0)}_\lambda(x,y)|$ tends to zero uniformly in $y$ when $\lambda>0$ and $r\to\infty.$

\begin{proposition}\label{tno}
Let conditions a),b) hold. Then for every sequence $\alpha_n\to+0,$ one can find a set of points $\{x_n\in X\}$ such that the operator
\begin{equation}\label{fullop}
H=H_0-\sum_{n=1}^\infty\alpha_n\delta_{x_n}(x)
\end{equation}
has infinitely many negative eigenvalues.
\end{proposition}
\textbf{Proof.} In order to prove the theorem, it is sufficient to construct a sequence of compactly supported functions $\{\psi_k(x)\}$ with disjoint supports such that
\begin{equation}\label{diri}
(H\psi_k(x),\psi_k(x))<0.
\end{equation}

For fixed $y\in X,\alpha>0$, consider a ``test" operator
$$
H=H(y,\alpha)=H_0-\alpha\delta_{y}(x).
$$
Due to Proposition \ref{tinter}, this operator has a negative eigenvalue $-\lambda_0(y,\alpha)$, where $\lambda=\lambda_0(y,\alpha)>0$ is the root of the equation  $-\alpha R^{(0)}_\lambda(y,y)=1$. The corresponding eigenfunction can be chosen as
$$
\psi(x)=\frac{R^{(0)}_{\lambda_0}(x,y)}{\sqrt{\sum_x[R^{(0)}_{\lambda_0}(x,y)]^2}}.
$$
Note that condition a) implies that $\lambda_0\geq \lambda_0(\alpha)>0$, where the lower bound $\lambda_0(\alpha)$ does not depend on $y$.

In order to complete the proof of the theorem, we will need the following lemma.
\begin{lemma}\label{unif}
There exists a function $r=r(\alpha)$ such that the inequality
\begin{equation} \label{un1}
(H\widetilde{\psi}(x),\widetilde{\psi}(x))<0,~~~H=H(y,\alpha),
\end{equation}
holds for the truncated eigenfunction
$$
\widetilde{\psi}(x)=\psi(x)I_{d(x,y)\leq r(\alpha)}.
$$
\end{lemma}

The important part of the statement of this lemma is that $r$ is $y-$independent. The statement follows from the uniformity assumption. Indeed,
$$
(H\psi(x),\psi(x))=-\lambda_0(y,\alpha)\leq-\lambda_0(\alpha)<0.
$$
Hence, it is enough to show that
$$
|(H\psi,\psi)-(H\widetilde{\psi},\widetilde{\psi})|=
|(H\psi,\psi-\widetilde{\psi})+(H(\psi-\widetilde{\psi}),\widetilde{\psi})|<\frac{\lambda_0(\alpha)}{2}\quad \text{when}~~r>r(\alpha).
$$
Since the operator $H$ is bounded in  $l^2(X)$, $\|\psi\|=1,~\|\widetilde{\psi}\|\leq 1$, it remains to prove that $\|\psi-\widetilde{\psi}\|\to 0$ uniformly in $y$ when $\alpha$ is fixed and $r\to\infty$ (all the norms here and below are in $l^2(X)$). It was shown in the proof of Theorem \ref{tinter} that $|R^{(0)}_\lambda(x,y)|\leq \frac{1}{\lambda}$ for each $x,y\in X$ and $\lambda>0$. Thus, from condition b) and the estimate $\lambda_0(y,\alpha)\geq \lambda_0(\alpha)>0$ it follows that
$$
\|(1-I_{d(x,y)\leq r})R^{(0)}_{\lambda_0}(x,y)\|^2\leq \frac{1}{\lambda_0}\sum_{x:d(x,y)>r}| R^{(0)}_{\lambda_0}(x,y)| \to 0
$$
uniformly in $y$ when $r\to\infty.$ Condition a) implies that
$$
\sum_x[R^{(0)}_{\lambda_0}(x,y)]^2\geq [R^{(0)}_{\lambda_0}(y,y)]^2\geq \beta(\lambda_0)>0.
$$
This completes the proof of the lemma since
$$
\|\psi-\widetilde{\psi}\|^2=
\frac{\|(1-I_{d(x,y)\leq r})R^{(0)}_{\lambda_0}(x,y)\|^2}{\sum_x[R^{(0)}_{\lambda_0}(x,y)]^2}.
$$

Let us complete the proof of the theorem.
We fix $\alpha_1$, calculate $r=r(\alpha_1)$, select an arbitrary point $x_1$ and chose $\psi_1(x)$ to be the truncated eigenfunction of the ``test" operator $H(x_1,\alpha_1)$. Other points $x_n,~n>1,$ will be chosen outside of the support of $\psi_1.$ Thus inequality (\ref{un1}) with the ``test" operator $H(x_1,\alpha_1)$ implies the same inequality for operator (\ref{fullop}), i.e., (\ref{diri}) holds for $\psi_1$. Now we fix $\alpha_2$, calculate $r=r(\alpha_2)$, select an arbitrary point $x_2$ such that $d(x_2,x_1)>r(\alpha_2)+r(\alpha_1)$, and chose $\psi_2(x)$ to be the truncated eigenfunction of the ``test" operator $H(x_2,\alpha_2)$. The supports of functions $\psi_1$ and $\psi_2$ are disjoint, and other points $x_n, ~n>2,$ will be chosen outside of the supports of $\psi_1,\psi_2$. Hence,  (\ref{diri}) holds for $\psi_2$. The point $x_3$ will be chosen in such a way that  $d(x_3,x_1)>r(\alpha_3)+r(\alpha_1)$ and $d(x_3,x_2)>r(\alpha_3)+r(\alpha_2)$, etc.. This procedure allows us to construct the desired sequence $\{\psi_k(x)\}$.
\qed

\section{Estimates from below} The goal of this section is to show that the estimates (\ref{barg}), (\ref{234}) are sharp in the following sense: the operator has infinitely many negative eigenvalues in the case of any potential which decays at infinity a little slower (by a logarithmic factor) than in those estimates. To be more exact, the following theorem holds
\begin{theorem}Let $H=-\Delta-V(x)$ be a one-dimensional Schr\"{o}dinger operator in $L^2(Z)$ or $L^2(R)$ with the potential $V$ such that for some $\varepsilon>0,$
\[
\sum_Z \frac{ |x|}{\ln^{1+\varepsilon}(1+|x|)} V(x)=\infty, \quad \text{or} \quad \int_R \frac{ |x|}{\ln^{1+\varepsilon}(1+|x|)}V(x)dx=\infty,
\]
respectively. Then $H$ has infinitely many negative eigenvalues ($N_0(V)=\infty$).

Let $H=-\Delta-V(x)$ be a two-dimensional Schr\"{o}dinger operator in $L^2(Z^2)$ and
\begin{equation} \label{6}
\sum_{Z^2}V(x)=\infty.
\end{equation}
Then $N_0(V)=\infty$.
\end{theorem}
\textbf{Proof.} We will prove the first statement ($d=1$) only in  the lattice case. The continuous case can be treated similarly (and, in fact, is simpler). Consider sets $l=l_k=\{x:2^k\leq x\leq 2^{k+1}\}\subset Z$. Let
\begin{equation} \label{3}
a_k=\sum_{l_k} \frac{ |x|}{\ln^{1+\varepsilon}(1+|x|)} V(x), \quad k
\geq 1.
\end{equation}
Since $\sum a_k=\infty$, there exists an infinite sequence of values of $k=k_j,~j=1,2,...,$ for which
\begin{equation} \label{4}
a_k>k^{-(1+\varepsilon /2)}, \quad k=k_j.
\end{equation}
By taking a subsequence, if needed, we can guarantee that $k_{j+1}-k_j\geq 2$. Let $L_k=\{x:2^{k-1}\leq x\leq 2^{k+2}\},~ k=k_j,$ be the union of $l_{k_j}$ and two neighboring sets $l_k$. The sets $\{L_{k_j}\}$ do not have common points (except, perhaps, the end points). The first statement of the theorem will be proved if, for infinitely many sets $L=L_{k_j}$, we construct functions $\psi=\psi_j$ with the support in $L$ and such that $(H\psi,\psi)<0$.

We will take
\[
\psi=\sin[\frac{\pi}{|L|}(x-a)],~~ x\in L, \quad \psi =0,~~ x \notin L,
\]
where $|L|=2^{k+2}-2^{k-1},~k=k_j,$ is the length of the interval between the end points of $L$ and $a=2^{k_j-1}$ is the left end point of the set $L$. The function $\psi$ is a sine function whose half-period is $L$ and which is zero outside $L$. The $l^2(Z)-$norm of this function for large $L$ has order $\sqrt{L/2}$:
\[
||\psi||=\sqrt{|L|/2}(1+o(1)), \quad |L|\to \infty.
\]
One can easily show that $-\Delta \sin\alpha x=\sigma (\alpha) \sin \alpha x,~x\in Z$,  where $\alpha $ is arbitrary and $\sigma (\alpha)=2-2\cos \alpha \sim \alpha^2$ as $\alpha \to 0$. Hence,
\[
-\Delta \psi =\sigma(\frac{\pi}{|L|}) \psi-\sin \frac{\pi}{|L|}(\delta _{a}(x)-\delta _{b}(x)),
\]
where $\delta _{y}(x)$ is the delta function at the point $y$, and $a,~b$ are the left and right end points of $L$, respectively. Thus, $(-\Delta \psi,\psi)=\sigma(\frac{\pi}{|L|}) ||\psi||^2$, and therefore
\begin{equation} \label{5}
(-\Delta \psi,\psi)\sim \frac{\pi^2}{2|L|}, \quad |L| \to \infty.
\end{equation}

Let us evaluate now
\[
(V\psi,\psi)=\sum_{x\in L}V(x)\psi^2(x)\geq \sum_{x\in l}V(x)\psi^2(x).
\]
Since $\frac{ |x|}{\ln^{1+\varepsilon}(1+|x|)}\leq C2^{k}k^{-1-\varepsilon}$ on $l_k$ and $V(x)\geq 0$, from (\ref{3}) and
(\ref{4}) it follows that
\[
\sum_{x\in l_k}V(x)\geq C_12^{-k}k^{\varepsilon/2}\geq C_1|L|^{-1}\ln^{\varepsilon/2}|L|, \quad |L|\to \infty.
\]
Furthermore, $l_k$ is located far enough from the end points of $L_k$, and there exists $c>0$ such that $\psi(x)>c,~ x\in l_k.$ Hence,
\[
(V\psi,\psi)\geq \frac{C\ln^{\varepsilon/2}|L|}{|L|}, \quad |L|\to \infty.
\]
Together with (\ref{5}), this proves that $(H\psi,\psi)\leq 0$ for large enough $L$.

The proof of the one-dimensional statement of Theorem \ref{t1} is complete.

Let us prove the statement of the theorem concerning the two-dimensional operators. As in the previous case, we will construct a sequence of functions $\psi=\psi_j(x),~x\in Z^2,$ with non-intersecting finite supports and such that $(H\psi,\psi)<0$. The functions $\psi_j$ will be defined by induction as the restrictions of some functions $\phi=\phi_j $ on the Euclidian space $R^2$ onto $Z^2 \subset R^2.$ Denote by $Q_k$ the squares in $R^2$ for which $|x_1|,|x_2|\leq k$. Let us define $\phi=\phi_{j_0+1}$ while assuming that the functions $\phi_j,~ j\leq j_0,$ have been constructed. We choose $k$ large enough so that the supports of all the functions $\psi_j$ already defined are located strictly inside $Q_k$. We take $k=1$ to define the first function $\phi_1$. The function $\phi=\phi_{j_0+1}$ will be supported on a square layer $P=Q_{2l}\setminus Q_{k} $ with some $l\gg k$ chosen below. Thus each layer $P_j$ is split naturally in two parts, the interior part $P^{(1)}=Q_{l}\setminus Q_{k}$ and the exterior part $P^{(2)}=Q_{2l}\setminus Q_{l}$. We put $\phi=0$ outside of $P$ and $\phi=1$ on the interior part of the layer $P$. Then we split the exterior part $P^{(2)}$ into four trapezoidal regions using diagonals of the square $Q_{2l}$ and define $\phi$ to be such a linear function in each of these trapezoidal regions that $\phi=1$ on the boundary $\partial Q_{l}$ of the square $Q_{l}$ and $\phi=0$ on $\partial Q_{2l} $. Note that $\phi=0 $ on $\partial P$.

Let us estimate $(-\Delta \psi, \psi),~ x\in Z^2,$ from above. We will use the notation $\partial Q_k$ for the boundary of the square $Q_k\subset R^2$, and $q$ for the union of the boundaries of the trapezoidal regions in $P \subset R^2$ constructed above. Since $-\Delta u=0$ for any linear function $u$ on $Z^2$, the support of the function $-\Delta \psi$ belongs to the set $\partial Q_{k}\bigcup\partial Q_{k+1}\bigcup q$, i.e.,
\[
|(-\Delta \psi, \psi)|\leq \sum_{\partial Q_{k}\bigcup\partial Q_{k+1}\bigcup q}|\Delta \psi|,
\]
since $0\leq\psi\leq 1$. Furthermore, $|\psi(x_1)-\psi(x_1)|\leq 1$ for each pair of neighboring points $x_1,x_2\in Z^2$, and therefore  $|-\Delta \psi|\leq 4,~x\in Q_{k}\bigcup\partial Q_{k_{j}+1}$. In fact, the latter estimate holds with $2$ instead of $4$, but we do not need this improvement. A better estimate holds on $q$. Since $|\nabla \phi|\leq 1/l$, we have $|-\Delta \psi|\leq 4/l,~x\in q$. Taking into account that $|\partial Q_{k}|+|\partial Q_{k}|\leq c_1k$ and $|l|\leq c_2l$, we arrive at
\[
|(-\Delta \psi, \psi)|\leq 4c_1k+4c_2.
\]
Note that the latter estimate does not depend on $l$.

Obviously, $(V\psi,\psi)\geq \sum _{x\in P^{(1)}}V(x)$. Assumption (\ref{6}) allows us to choose $l$ such that the right-hand side of the latter inequality exceeds $4c_1k+4c_2$. Then $(H\psi,\psi)<0$ and the proof is complete.
\qed

\section{Fractional power of the lattice operator}

This section provides an illustration of the results on general discrete operators obtained in Theorem \ref{t1vv}. It concerns an important class of non-local random walks  $x_\alpha(t)$ on the one-dimensional lattice. The one-dimensional lattice Laplacian
$$
-H_0 =\Delta\psi(x)=\psi(x+1)+\psi(x-1)-2\psi(x)
$$
on $l^2(Z)$ is the generator of the symmetric random walk $x(t)$ with continuous time. Let $P_t=e^{t\Delta}$ be the corresponding semigroup, $P_t\psi=\sum_{y\in Z}p(t,x,y)\psi(y)$. The operator
$\Delta$ in the Fourier space $L^2[-\pi,\pi]$ acts as multiplication by the symbol
$$
\widehat{\Delta}(\phi)=2(\cos\phi-1)=-4\sin^2\frac{\phi}{2},~~\phi\in[-\pi,\pi].
$$
Similarly,
$$
\widehat{P_t}(\phi)=e^{-4t\sin^2\frac{\phi}{2}}, \quad \widehat{R}_\lambda(\phi)=-\int_0^\infty e^{-\lambda t}\widehat{P_t}dt=\frac{-1}{\lambda+4\sin^2\frac{\phi}{2}},~~\phi\in[-\pi,\pi].
$$

The main objects that we study in this section are the fractional degrees $H_0^\alpha, ~\alpha > 0,$ of the operator $H_0=-\Delta$. After the Fourier transform, the operator $H_0^\alpha$ , its semigroup and the resolvent are the operators of multiplication by the symbols
$$
\widehat{(-\Delta)^\alpha}=\left(4\sin^2\frac{\phi}{2}\right)^\alpha, \quad \widehat{P_{t,\alpha}}=e^{-t\left(4\sin^2\frac{\phi}{2}\right)^\alpha}, \quad
\widehat{R_{\lambda,\alpha}}=\frac{-1}{\lambda+(4\sin^2\frac{\phi}{2})^\alpha}.
$$

The following result is well-known in probability theory.
\begin{lemma}\label{polo}
The operator $-H_0^\alpha=-(-\Delta)^\alpha,~\alpha>0,$ is the generator of a Markov process $x_\alpha(t)$ on $Z$ iff $0< \alpha \leq 1$.
\end{lemma}
\textbf{Proof.} One needs only to prove the positivity of the kernel $p_\alpha (t,x,y)$ of the semigroup $P_{t,\alpha}$ for $0<\alpha<1$ and non-positivity of the kernel for $\alpha>1$. If $0<\alpha< 1$, then there exists \cite{F} (Ch. 13, 6) the probability density $g_{\alpha, 1}(s)>0,~0< s<\infty,$ (which is called the stable law with the parameters $\alpha$ and $\beta=1$) such that
$$
e^{-\lambda^\alpha}=\int_0^\infty e^{-\lambda t}g_{\alpha, 1}(t)dt.
$$
The second parameter $\beta$ in the two-parametric family of the densities $g_{\alpha, \beta}$ characterizes the symmetry of the density. If $\beta=0$ then $g_{\alpha, 0}(s)= g_{\alpha, 0}(-s)$, if $\beta=1,~0< \alpha <1,$ then $g_{\alpha, 1}(s)=0,~s\leq 0,~ g_{\alpha, 1}(s)>0,~s>0$.

Using the probability density $g_{\alpha, 1}$, one can represent $P_{t,\alpha}$ in the form
$$
P_{t,\alpha}=e^{-t(-\Delta)^\alpha}=\int_0^\infty e^{t^{1/\alpha} s\Delta}g_{\alpha, 1}(s)ds=\int_0^\infty P_{t^{1/\alpha}s}g_{\alpha, 1}(s)ds, \quad 0<\alpha<1,
$$
i.e., the kernels $p_\alpha$ and $p$ of the operators $P_{t,\alpha},~P_t$ are related by
$$
p_\alpha(t,x,y)=\int_0^\infty p(t^{1/\alpha} s,x,y)g_{\alpha, 1}(s)ds, \quad 0<\alpha<1.
$$
This implies the positivity of $p_\alpha$ when $0<\alpha<1$.

In order to show that the semigroup $P_{t,\alpha}$ is not positive when $\alpha>1$, we note that the function
$
\widehat{h}(t,\phi)=\widehat{P_{t,\alpha}}(\phi)
$ has the following property: $\widehat{h}''(0)=0$. Hence, its inverse Fourier transform $h(t,z)=\frac{1}{2\pi}\int_{-\pi}^\pi \widehat{h}(t,\phi)e^{iz\phi}d\phi$ satisfies $\sum_{z\in Z}z^2h(t,z)=0$, which shows that $p_\alpha=h(t,x-y),~\alpha>1,$ can not be non-negative.
\qed

\begin{lemma}\label{polo}
For each $\alpha\in (0,1]$ and $t\to\infty$,
$$
p_\alpha(t,x,x)\sim \frac{c_\alpha}{t^{1/(2\alpha)}},  \quad  c_\alpha=\frac{\Gamma(1/(2\alpha))}{2\pi\alpha}.
$$
\end{lemma}
\textbf {Corollary.} The random walk $x_\alpha(t)$ is transient for $0<\alpha<1/2$ and recurrent for $1/2\leq\alpha\leq 1.$ The formula above indicates that $p_\alpha(t,x,x)$ has the same asymptotic behavior as the transition probability for ``nearest neighbors random walks" (defined by the standard Laplacian) when the dimension $d$ equals $1/\alpha$. A similarity between the long range 1-D ferromagnetic interaction and high-dimensional local interaction was discovered by Dyson \cite{Dy}. This similarity was a foundation for the introduction of the hierarchical lattice which will be discussed in the last section.

\textbf {Proof.} This statement is a simple consequence of the Laplace method applied to the integral
$$
p_\alpha(t,x,x)=\frac{1}{2\pi}\int_{-\pi}^\pi e^{-t\left(4\sin^2\frac{\phi}{2}\right)^\alpha}d\phi\sim \frac{1}{2\pi}\int_{-\pi}^\pi e^{-t|\phi|^{2\alpha}}d\phi, \quad t\to\infty.
$$
\qed

\textbf{Remark.} Similar calculations give a more general result. If $x_\alpha(t)$ is a random walk on $Z$ with the generator $-(-\Delta)^\alpha$, then
\begin{equation} \label{law}
\frac{x_\alpha(t)}{t^{1/\alpha}}\rightarrow \phi \quad \text{in law as} ~~t\to\infty,
\end{equation}
where $\phi$ has a stable distribution $g_{2\alpha,0}(s),~s\in R,$ (symmetric stable law with parameters $2\alpha, \beta=0$ and characteristic function (Fourier transform) equal to $e^{-\lambda^{2\alpha}}$). Indeed,
$$
E_0e^{i\frac{\lambda x_\alpha(t)}{t^{1/\alpha}}}=e^{-t\left(4\sin^2\frac{\lambda}{t^{1/\alpha}}\right)^\alpha}\to e^{-\lambda^{2\alpha}},~~~t\to\infty.
$$
Formula (\ref{law}) means that, after rescaling, the lattice operator $(-\Delta)^\alpha$ approximates the fractional power of the continuous Laplacian, i.e., the random walk $x'_\alpha(s)=x_\alpha(st)/t^{1/\alpha}$ (which is the rescaling of $x_\alpha(s)$) converges to the stable process $x^*_\alpha(s)$ on $R$ with the generator $-(-\frac{d^2}{dx^2})^\alpha$.

The following theorem is an immediate consequence of the standard CLR estimate (\ref{1xx}) with $\sigma=0$, Lemma \ref{polo}, and finite rank perturbation arguments (see Remark 2 after Theorem \ref{t1})

\begin{theorem} \label{576}(Transient case) Consider the Hamiltonian
$$
H_\alpha=-(-\Delta)^\alpha-V(x) \quad \text{on} ~~l^2(Z),~~~V(x)\geq 0,
$$
with $0<\alpha<1/2$ (i.e., the dimension $d=\frac{1}{\alpha}>2$). Then there is a constant $C=C(\alpha)$ such that
$$
N_0(V)\leq \sum_{x\in Z}\min (1, ~C(\alpha)V^{\frac{1}{2\alpha}}(x)).
$$
\end{theorem}
\textbf{Remark.} The constant $C(\alpha)$ can be evaluated. One can show that $C(\alpha)=O(\frac{1}{1-2\alpha})$ as $\alpha\to 1/2.$

Let us consider now the recurrent case: $\alpha\geq 1/2.$ First we calculate the regularized resolvent (\ref{res}):
$$
\widetilde{R}_0(x,0)=2\lim_{\lambda\to +0}[R_{\lambda,\alpha}(x,0)-R_{\lambda,\alpha}(0,0)]=\lim_{\lambda\to +0}
\frac{2}{\pi}\int_{-\pi}^\pi \frac{1-e^{i\phi x}}{\lambda+(4\sin^2\frac{\phi}{2})^\alpha}d\phi
$$
$$
=\lim_{\lambda\to +0}\frac{4}{\pi}\int_{0}^\pi \frac{\sin^2(\frac{\phi}{2}x)}{\lambda+(4\sin^2\frac{\phi}{2})^\alpha}d\phi
=\frac{4}{\pi}\int_{0}^\pi \frac{\sin^2(\frac{\phi}{2}x)}{(4\sin^2\frac{\phi}{2})^\alpha}d\phi.
$$

A simple analysis provides the following estimates:
$$
\widetilde{R}_0(x,0)\leq C\frac{(1+|x|)^{2\alpha-1}-1}{2\alpha-1},\quad 1/2<\alpha\leq 1; \quad  \widetilde{R}_0(x,0)\leq C\ln (1+|x|), \quad \alpha= 1/2.
$$

Hence, Theorem \ref{barggen} implies
\begin{theorem} (Recurrent process)
There exist constant $C$ such that
$$
N_0(V)\leq 1+\sum_{x\in Z}\min (1,~CV(x)\frac{(1+|x|)^{2\alpha-1}-1}{2\alpha-1}), \quad ~~\frac{1}{2}<\alpha<1,
$$
$$
N_0(V)\leq 1+\sum_{x\in Z}\min (1,~C V(x)\ln(1+|x|)), \quad ~~\alpha=1/2.
$$
\end{theorem}
\textbf{Remark.} The first estimate remains valid for $\alpha<1/2$ (the transient case) and provides a better result than Theorem \ref{576} when $\alpha$ is close to $1/2.$

\section{Lieb-Thirring sums}

The results of this section are based on two known formulas for the Lieb-Thirring sums for the general  Schr\"{o}dinger operators $H=H_0-V(x)$ on a complete $\sigma$-compact metric space $X$. The first formula is valid under the same assumptions that are needed for formula (\ref{1xx}) and has the form
\begin{equation} \label{lit}
S_\gamma(V)\leq\frac{1}{c(\sigma)}\int_X V^{1+\gamma}(x)\int_{\frac{\sigma}{V(x)}}^\infty p_0(t,x,x)dt\mu(dx).
\end{equation}
Note that the well-known estimate
\begin{equation} \label{lith1}
S_\gamma(V)\leq c_{d,\gamma}\int_{R^d}V^{\frac{d}{2}+\gamma}dx,~~\frac{d}{2}+\gamma>1
\end{equation}
for the operator  $H=-\Delta-V(x)$ in $R^d$ is an immediate consequence of (\ref{lit}) only if $d\geq 3$ (formula (\ref{lit}) is meaningless if the underlying Markov process is recurrent). The next formula for $S_\gamma(V)$ is valid under the same conditions which are needed for  (\ref{1xx}), but the transience requirement is replaced by the convergence of the following integral:
$$
\int_1^\infty t^{-\gamma}p_0(t,x,x)dt<\infty.
$$
If the latter integral converges, then
\begin{equation} \label{lithi}
S_\gamma(V)\leq\frac{2\gamma \Gamma(\gamma)}{c(\sigma)}\int_X V(x)\int_{\frac{\sigma}{V(x)}}^\infty t^{-\gamma}p_0(t,x,x)dt\mu(dx),
\end{equation}
where $\Gamma(\gamma)$ is the Gamma-function. Note that (\ref{lithi}) implies (\ref{lith1}) for the operator  $H=-\Delta-V(x)$ in $R^d$ when $\frac{d}{2}+\gamma>1,$ i.e., the case $d=1,~ \gamma\leq 1/2$  is still not covered by (\ref{lit}),(\ref{lithi}). The important borderline case $d=1,~\gamma=1/2$ can be found in \cite{dirk}, \cite{hs}. Estimates for other cases are obtained below. Our approach allows us also to obtain a uniform in $\gamma$ estimate which is valid for all $d,\gamma$ and is better than (\ref{lith1}) in the case $\gamma=1-\frac{d}{2}+\varepsilon,~ \varepsilon\to+0 $. Note that $c_{d,\gamma}\to \infty$ as $\varepsilon\to +0$.

 While estimate (\ref{lit}) can be found in many papers starting from the original paper by Lieb and Thirring \cite{Lt1} (see also \cite {rs}, \cite{1}, \cite{mv}), we didn't find a reference for  (\ref{lithi}) (it is given as an exercise in \cite{rs}). Thus we decided to outline the proof of (\ref{lithi}). Let $N_E(V)=\#\{\lambda_i\leq -E,E>0\}$. Assume first that $\int_0^1 tp_0(t,x,x)dt<\infty$ (for example, $H_0=-\Delta$ in $R^3$). Then we have
$$
S_\gamma(V)=\gamma\int_0^\infty E^{\gamma-1}N_E(V)dE\leq 2\gamma \int_0^\infty E^{\gamma-1}\text{Tr}(V[(H_0+E)^{-1}-(H_0+V+E)^{-1}])dE
$$
$$
\leq 2\gamma \int_0^\infty E^{\gamma-1}\text{Tr}(V\int_0^\infty[e^{-t(H_0+E)}-e^{-t(H_0+V+E)}]dt)dE
$$
$$
=2\gamma \Gamma(\gamma)\int_0^\infty t^{-\gamma}\text{Tr}(V[e^{-tH_0}-e^{-t(H_0+V)}])dt.
$$
An estimate of the trace in the right-hand side above based the Kac-Feynman formula implies (\ref{lithi}), see details in \cite{rs}, proof of Theorem XIII.12. If $p_0$ has a stronger singularity at $t=0$, one needs to use the inequality \cite{rs}
$$
N_E(V)\leq (m+1)\textrm{Tr}V\sum_0^m (-1)^j\left (
\begin{array}{cc}
j \\
m  \\
\end{array} \right )(H_0+jV+E)^{-1}
$$

Consider now a Schr\"{o}dinger operator $H=H_0-V(x)$ on a metric space $X$ such that the Markov process $x(t)$, generated by $-H_0$, is recurrent, and a point $x_0$
is accessible from any initial point. Then the process $x_1(t)$ with annihilation at the moment of the first arrival to $x_0$ is transient. For example, $H_0$ can be a negative lattice Laplacian on $Z^d,d\leq 2,$ the general discrete operator discussed in section 6, or the generator of a 1-D diffusion process, say, $H_0=-\frac{d^2}{dx^2}, ~x\in R$. Let $-H_1$ be the generator of the process $x_1(t)$. It is given by $-H_0$ with the Dirichlet boundary condition at $x_0:~\psi(x_0)=0$. Let $p_1(t,x,y)$ be the transition probability for the process $x_1(t)$.

We will assume additionally that the potential is bounded: $V(x)\leq \Lambda.$ This implies that the ground state $\lambda_0(V)$ is bounded from below, $\lambda_0(V)\geq -\Lambda$. Since the operator $\widetilde{H}=H_1-V(x)$ is a rank one perturbation of $H=H_0-V(x)$, the eigenvalues of the operators $\widetilde{H}$ and $H$ alternate. Hence, the bound for the ground state and estimates (\ref{lit}),(\ref{lithi}) applied to the operator $\widetilde{H}$ lead to the following statement.
\begin{theorem}\label{tlt}
Let $S_\gamma(V)$ be the Lieb-Thirring sum for the Schr\"{o}dinger operator $H=H_0-V(x)$, where $H_0$ is an operator which satisfies the conditions described above (including the recurrence and the accessibility of some point $x_0$), and $V(x)\leq \Lambda$. Then the following two estimates hold for each $\sigma\geq 0$:
\begin{equation} \label{lit9}
S_\gamma(V)\leq\Lambda^\gamma+\frac{1}{c(\sigma)}\int_X V^{1+\gamma}(x)\int_{\frac{\sigma}{V(x)}}^\infty p_1(t,x,x)dt\mu(dx),
\end{equation}
\begin{equation} \label{lithi9}
S_\gamma(V)\leq\Lambda^\gamma+\frac{2\gamma \Gamma(\gamma)}{c(\sigma)}\int_X V(x)\int_{\frac{\sigma}{V(x)}}^\infty t^{-\gamma}p_1(t,x,x)dt\mu(dx).
\end{equation}
In particular, if $\sigma=0$, then
\begin{equation} \label{lit9a}
S_\gamma(V)\leq\Lambda^\gamma+\int_X V^{1+\gamma}(x)\widetilde{R}(x,x_0)\mu(dx),
\end{equation}
where $\widetilde{R}$ is defined by (\ref{res}), and
\begin{equation} \label{lithi9a}
S_\gamma(V)\leq\Lambda^\gamma+2\gamma \Gamma(\gamma)\int_X V(x)\int_{0}^\infty t^{-\gamma}p_1(t,x,x)dt\mu(dx).
\end{equation}
\end{theorem}
\textbf{Remark.} The estimate (\ref{lithi9}) can be applied to the potentials which decay at infinity slower than in (\ref{lit9}), but it worsens when $\gamma\to\gamma_0$ and $t^{-\gamma_0}p_1$
 is not integrable at zero.

Estimates (\ref{lit9a}), (\ref{lithi9a}) are Bargmann type estimates, and (\ref{lit9}), (\ref{lithi9}) are refined Bargmann type estimates.

 Let us apply  (\ref{lit9a}) and (\ref{lithi9a}) to the one- and two-dimensional Schr\"{o}dinger operators.

\textbf{Example 1.} Let $H=-\frac{d^2}{dx^2}-V(x)$ in  $L^2(R)$ and $0\leq V(x)\leq \Lambda$. We choose $x_0=0$ and construct the transition density of the Brownian motion with the killing at  $x_0=0$. Then (see Example 1 in section 3):
$$
p_1(t,x,x)=\frac{1-e^{-\frac{x^2}{t}}}{\sqrt{4\pi t}}, \quad ~\widetilde{R}(x,x_0)=\int_{0}^\infty p_1(t,x,x)dt=|x|.
$$
Hence (\ref{lit9a}) implies
\begin{equation}\label{987}
S_\gamma(V)\leq\Lambda^\gamma+\int_{-\infty}^\infty V^{1+\gamma}(x)|x|dx.
\end{equation}

The latter inequality with $\gamma=0$ becomes the usual Bargmann estimate for $N_0(V)$. If $\gamma=1/2$, inequality (\ref{987}) is worse than a very fine estimate obtained in \cite{dirk}:
\begin{equation}\label{drk}
\sum|\lambda_i|^{1/2}\leq\frac{1}{2}\int_{-\infty}^\infty Vdx.
\end{equation}

Now let us apply (\ref{lithi9a}) to our particular operator $H$. Note that, for $ \gamma<1/2$,
$$
\int_{0}^\infty t^{-\gamma}p_1(t,x,x)dt=\int_{0}^\infty \frac{1-e^{-\frac{x^2}{t}}}{2\sqrt\pi t^{1/2+\gamma}}dt=
\frac{|x|^{1-2\gamma}}{2\sqrt\pi}\int_{0}^\infty \frac{1-e^{-1/s}}{s^{1/2+\gamma}}ds=c(\gamma)|x|^{1-2\gamma},
$$
i.e.,
\begin{equation*}
S_\gamma(V)\leq\Lambda^\gamma+2\gamma \Gamma(\gamma)c(\gamma)\int_X V(x)|x|^{1-2\gamma}dx.
\end{equation*}
This estimate also coincides with the Bargmann estimate for $N_0(V)$ when $\gamma\to 0$ and is worse than (\ref{drk}) when $\gamma\to 1/2$ since the coefficient at the integral explodes as $\gamma \to 1/2$.

\textbf{Example 2.} Consider the two-dimensional lattice Schr\"{o}dinger operator $H=-\Delta-V(x),~x\in Z^2$. The classical estimate (\ref{lithi}) with $\sigma =0, \gamma\in (0,1)$ implies
$$
S_\gamma(V)\leq 2\gamma \Gamma(\gamma)c_1(\gamma)\sum_{Z^2}V(x),\quad c_1(\gamma)=\int_0^\infty t^{-\gamma}p_0(t,0,0)dt.
$$
Since $p_0(t,0,0)$ is bounded and $p_0(t,0,0)\sim c/t,~t\to\infty$, we have $c_1(\gamma)=O(1/\gamma),~\gamma\to +0$, i.e., the right-hand side above explodes when $\gamma\to +0$.

The Bargmann estimate  (\ref{lithi9a}) leads to the following result. From Lemma \ref{ldiscrt} it follows that
\[
\int_0^\infty t^{-\gamma}p_1(t,x,x)dt\leq C\ln^2(2+|x|)
\]
uniformly in $\gamma\in (0,1-\varepsilon), ~\varepsilon>0,$ and therefore
\[
S_\gamma(V)\leq \Lambda^\gamma +C\gamma \Gamma(\gamma)\sum_{Z^2}V(x)\ln^2(2+|x|),\quad 0<\gamma<1-\varepsilon.
\]
This inequality has a regular behavior when $\gamma\in (0,1-\varepsilon)$ and allows one to pass to the limit as $\gamma \to 0$. The refined Bargmann estimate provides a somewhat better result.

\section{The hierarchical lattice}
In this section we apply the general Theorem \ref{nt2} on annihilation in the Markov chains to study the hierarchical
Laplacian. The concept of the hierarchical lattice was introduced by F. Dyson \cite{Dy} in a completely different setting (phase transition for 1-D ferromagnetic spin systems with a long range interaction). Some properties of the hierarchical Laplacian were analyzed in \cite {md} in connection with the random Anderson model. We will study the negative spectrum of the hierarchical Schr\"{o}dinger operator. Our results can be extended to the general nested fractals (similar to the infinite Sierpinski gasket or lattice) due to deep similarities between the spectral properties of the Laplacians on the hierarchical lattice and on the classical fractals. The most interesting part below concerns the case when the spectral dimension goes through the critical value $d=2$.

Consider a countable set $X$ and a family of partitions $\Pi_0\subset\Pi_1\subset\Pi_2\subset...$ of $X$. The elements $Q_i^{(0)}$ of $\Pi_0$ (cubes of rank zero) are the points of $X$. Each element $Q_i^{(1)}$ of $\Pi_1$ (cube of rank one) is a union of $\nu$ different cubes of rank zero, i.e., $X=\cup Q_i^{(1)},~| Q_i^{(1)}|=\nu$ (see Figure 1). Each element $Q_i^{(2)}$ of $\Pi_2$ (cube of rank two) is a union of $\nu$ different cubes of rank one, i.e., $X=\cup Q_i^{(2)},~| Q_i^{(1)}|=\nu^2$, and so on. The parameter $\nu\geq 2$ is one of the two basic parameters of the model.

\begin{figure}[!ht]
\vspace{0.3 cm}
\centering
\includegraphics[width=.6\textwidth]{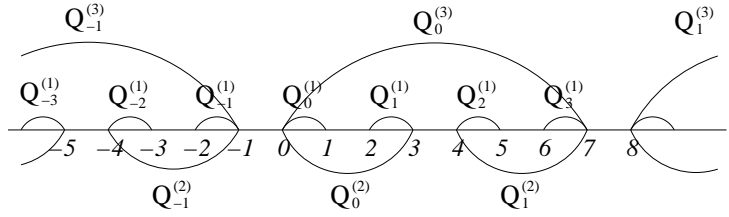}
\caption{An example of a hierarchical lattice with $X=l^1$ and $\nu=2.$}
\label{Picture1}
\end{figure}

Each point $x$ belongs to a sequence of increasing cubes of rank $r\geq 0$ which we denote by $Q^{(r)}(x)$, i.e., $x=Q^{(0)}(x)\subset Q^{(1)}(x)\subset Q^{(2)}(x)\subset...~$. The hierarchical distance $d_h(x,y)$ on $X$ is defined as follows
$$
d_h(x,y)=\min\{r:\exists Q^{(r)}_i\ni x,y\}.
$$
We assume that the following connectivity condition holds: for each $x,y\in X$, the cubes $Q^{(n)}(x)$ contain $y$ when $n$ is large enough, i.e., $d_h(x,y)<\infty.$

We denote by $l^2(X)$ the standard Hilbert space on the metric space $(X,d_h)$, and define the (hierarchical) Laplacian which depends on a parameter $p\in(0,1)$:
$$
\Delta \psi=\sum_{r=1}^\infty a_r[\frac{\sum_{x'\in Q^{(r)}(x)}\psi(x')}{\nu^r}-\psi(x)],\quad \text{where} \quad a_r=(1-p)p^{r-1}, \quad \sum_{r=1}^\infty a_r=1.
$$
The random walk on $(X,d_h)$ related to the hierarchical Laplacian has a simple structure. It spends an exponentially distributed time $\tau$ (with parameter one) at each site $x$. At the moment $\tau +0$ it jumps at one of the points $x'\in Q^{(k)}(x),~k\geq 1,$ where $k$ is selected randomly with $P\{k=r\}=a_r$ and with the position in $Q^{(k)}(x)$ being uniformly distributed.

It is clear that $\Delta=\Delta^*,~\Delta\leq 0,$ Sp$(\Delta)\in[-1,0]$. The spectral analysis of the hierarchical Laplacian is simple. The following proposition holds.
\begin{proposition}\label{prd1}
(a) The spectrum of $-\Delta$ consists of discrete eigenvalues $\lambda_k=p^{k-1},~k=1,2,...,$ of infinite multiplicity.

(b) The corresponding eigenspaces $L_k \subset l^2(X)$ have the following structure: if $k=1$, then
$$
L_1=\{\psi\in l^2(X):~\sum_{x\in Q_i^{(1)}}\psi(x)=0 \quad {\textrm{for each}} \quad Q_i^{(1)}\subset \Pi_1 \}.
$$
Space $L_k,~k>1,$ consists of $\psi\in l^2(X)$ which are constant on each cube $Q_{i}^{(k-1)}$, and $\sum_{x\in Q_i^{(k)}}\psi(x)=0$ for each $Q_i^{(k)}\subset \Pi_k$.

(c) For each $y\in X$,
$$
\delta_y(x)=(\delta_y(x)-\frac{I_{Q^{(1)}(y)}(x)}{\nu})+(\frac{I_{Q^{(1)}(y)}(x)}{\nu}-\frac{I_{Q^{(2)}(y)}(x)}{\nu^2})+
(\frac{I_{Q^{(2)}(y)}(x)}{\nu^2}-\frac{I_{Q^{(3)}(y)}(x)}{\nu^3})+...~~,
$$
where the first term belongs to $L_1$, the second term belongs to $L_2$, etc.

(d) The following decomposition holds:   $l^2(X)=\oplus_{r=0}^\infty L_r.$
\end{proposition}
\textbf{Proof.} One can show by direct inspection that $L_k$ consists of eigenfunctions with the eigenvalue $\lambda_k=p^{k-1}$, and that (c) holds. Statement (c) immediately implies (d) which justifies (a).
\qed

Proposition \ref{prd1} and the Fourier method lead to the following result. Let $p(t,x,y)$ be the transition function of the hierarchical random walk $x(t)$, i.e.,
$$
\frac{\partial p}{\partial t}=\Delta p, ~~p(0,x,y)=\delta_y(x).
$$
\begin{proposition}\label{prd2}
 The following expansion is valid for the transition function $p$:
$$
p(t,x,y)=-\frac{e^{-p^{r-1}t}}{\nu^r}-(\frac{1}{\nu^r}-\frac{1}{\nu^{(r+1)}})e^{-p^{r}t}-
(\frac{1}{\nu^{r+1}}-\frac{1}{\nu^{(r+2)}})e^{-p^{r+1}t}-...~~,
$$
where $r=d_h(x,y)$. In particular, for each $x\in X$,
$$
p(t,x,x)=(1-\frac{1}{\nu})[e^{-t}+\frac{e^{-pt}}{\nu}+\frac{e^{-p^2t}}{\nu^2}+\frac{e^{-p^3t}}{\nu^3}+...]~~.
$$
\end{proposition}
Integration in $t$ leads to the following expansions for the resolvent
$$
R_\lambda (x,y)=-\int_0^\infty e^{-\lambda t}p(t,x,y)dt,~~\lambda >0.
$$
\begin{proposition}\label{prd3}
The following expansions hold when $\lambda>0$:
$$
-R_\lambda (x,x)=(1-\frac{1}{\nu})[\frac{1}{1+\lambda}+\frac{1}{\nu (p+\lambda)}+...+\frac{1}{\nu ^s(p^s+\lambda)}+...]~~,
$$
$$
-R_\lambda (x,y)=-\frac{1}{\nu ^r(p^{r-1}+\lambda)}+(\frac{1}{\nu^r}-\frac{1}{\nu^{(r+1)}})\frac{1}{(p^r+\lambda)}+...~~,~~r=d_h(x,y),
$$
$$
R_\lambda (x,y)-R_\lambda (x,x)=(1-\frac{1}{\nu})[\frac{1}{1+\lambda}+\frac{1}{\nu (p+\lambda)}+...+\frac{1}{\nu ^{r-2}(p^{r-2}+\lambda)}]+\frac{1}{\nu ^{r-1}(p^{r-1}+\lambda)}.
$$
\end{proposition}
\begin{corollary} \label{dc1}
$$
\widetilde{R}(x,y)=2\lim_{\lambda\to+0}[R_\lambda (x,y)-R_\lambda (x,x)]=2(1-\frac{1}{\nu})[1+\frac{1}{\nu p}+...+\frac{1}{\nu ^{r-2}p^{r-2}}]+\frac{1}{\nu ^{r-1}p^{r-1}},
$$
$$
-R_0(x,x)=-\lim_{\lambda\to+0}R_\lambda (x,x)=(1-\frac{1}{\nu})\sum_{s=0}^\infty\frac{1}{(\nu p)^s}.
$$
The process $x(t)$ with the transition probability $p(t,x,y)$ is transient if $\nu p>1$ and recurrent if $\nu p\leq 1$.
\end{corollary}
\begin{definition}\label{defsh}
We will call the number $s_h=\frac{2\ln\nu}{\ln(1/p)}$ the spectral dimension of the hierarchical Laplacian $\Delta$.
\end{definition}
The following proposition justifies the definition.
\begin{proposition}\label{prd4}
For each $\nu, p$,
$$
p(t,x,x)\asymp \frac{1}{t^{s_h/2}}.
$$
To be more exact, there exists a positive periodic function $h(x)\equiv h(x+1)$ such that
$$
p(t,x,x)=\frac{h(\ln t/\ln(1/p))}{t^{s_h/2}}(1+o(1)) \quad \text{as}~~t\to\infty.
$$
\end{proposition}
\textbf{Remark 1.} The proposition provides an alternative proof of the transience of the random walk $x(t)$ with the transition probability $p(t,x,y)$ if $\nu p>1$ and its recurrency if $\nu p\leq 1$.

\textbf{Remark 2.} The presence of a logarithmically periodic oscillation in the transition probability is a common feature of all the ``classical" fractals similar to the Sierpinski lattice.

\textbf{Proof.} Denote by $[z]$ and $\{z\}$ the integer and fractional parts of a number $z\in R$. The maximal term in the series $p(t,x,x)=(1-\frac{1}{\nu})\sum_{s=0}^\infty\frac{e^{-p^st}}{\nu^s}$ has order $s=O(\frac{\ln t}{\ln(1/p)})$, $t\to\infty$. We put $k=[\frac{\ln t}{\ln(1/p)}]$ and change the order of terms in the series representation of $p$, first taking the sum  over $s\geq k$ and then taking the sum over $s<k$:
$$
p(t,x,x)=(1-\frac{1}{\nu})[\frac{e^{-p^{kt}}}{\nu^k}+\frac{e^{-p^{(k+1)t}}}{\nu^{k+1}}+...
+\frac{e^{-p^{(k-1)t}}}{\nu^{k-1}}+...]
$$
\begin{equation}\label{pper}
=(1-\frac{1}{\nu})\frac{e^{-p^{kt}}}{\nu^k}[1+\frac{e^{-p^{kt(1-p)}}}{\nu}+
\frac{e^{-p^{kt(1-p^2)}}}{\nu^2}+...+\frac{e^{-p^{kt(1-p^{-1})}}}{\nu^{-1}}+
\frac{e^{-p^{kt(1-p^{-2})}}}{\nu^{-2}}+...].
\end{equation}
Since $\frac{\ln t}{\ln(1/p)}=k+\{\frac{\ln t}{\ln(1/p)}\}$, we have
$$
p^{kt}=e^{-\{\frac{\ln t}{\ln(1/p)}\}} \quad \text{and} \quad \frac{1}{\nu^k}=e^{-\frac{\ln t}{\ln(1/p)}\ln \nu} \nu^{-\{\frac{\ln t}{\ln(1/p)}\}}=\frac{1}{t^{s_h/2}}\nu^{-\{\frac{\ln t}{\ln(1/p)}\}}.
$$
We substitute the latter relations into (\ref{pper}) and note that $\{x\}$ is a periodic function of $x$ with period one. Now the proof can be easily completed.
\qed

We are going now to apply the standard CLR estimate (\ref{1xx}) and our results on the counting function $N_0(V)$  of the Schr\"{o}dinger operator $H=-\Delta-V(x),~V\geq 0,$ where $\Delta $ is the hierarchical Laplacian. Proposition \ref{prd4} implies that CLR estimate (\ref{1xx}) is meaningful only in the transient case $s_h>2$ and that the constant in the estimate is of order $O(\frac{1}{s_h-2})$ as $s_h\to 2+0$. To be more exact, the following theorem is a consequence of (\ref{1xx}):
\begin{theorem}
If $s_h>2$, then
$$
N_0(V)\leq \#\{x:~V(x)\geq 1\}+C(p,\nu)\sum_{x:V(x)<1}V^{s_h/2}(x),
$$
where $C(p,\nu)\to\infty$ as $s_h\downarrow 2$.
\end{theorem}
When $s_h=2+\varepsilon,~\varepsilon>0$, the estimate above is not valid with a constant independent of $\varepsilon$, since the operator $-\Delta-\sigma\delta_{x_0}(x)$ has an eigenvalue when $\sigma=O(s_h-2)$. Our results from section 3 allow one to obtain an estimate on $N_0(V)$ for all $s_h$. Besides, the constant in the estimate is uniformly bounded as  $s_h\to 2$. Hence these results are useful not only when $s_h\leq 2$, but also for $s_h=2+\varepsilon,~\varepsilon>0$.

We will apply now Theorem \ref{t1vv} to the hierarchial Schr\"{o}dinger operator $H=-\Delta-V(x),~V\geq 0,$ but first we introduce a different metric on $X$. Fix a point $x_0\in X$ (the origin) and put
$$
\rho(x,y)=p^{-\frac{1}{2}\max(d_h(x_0,x),d_h(x_0,y))}-1.
$$
Since different cubes $Q_i^{(r)}$ of the same rank $r$ do not have common points, it follows that $d_h(x,y)\leq\max_z(d_h(x,z),d_h(y,z))$, and the latter property implies that $\rho$ is a metric. This metric is somewhat closer to the usual distance on $R$. In particular, from the series representation of $R_0$ (Corollary \ref{dc1}) it follows that
$$
-R_0(x,y)\asymp \rho(x,y)^{2-s_h}\quad \text{when}\quad s_h>2.
$$
Similar calculations show that
$$
\widetilde{R}_0(x_0,x)\asymp \rho(x,y)^{2-s_h},\quad s_h<2;
$$
$$
\widetilde{R}_0(x_0,x)\asymp
d_h(x_0,x)=\frac{-2\ln\rho}{\ln p},\quad  s_h=2~~(\text{i.e.,}~~\nu p=1).
$$

The following uniform in $s_h\in [1,2]$ estimates are valid.
\begin{proposition}\label{prunif}
There exists a constant $c_0$ which is independent of $s_h\in[1,2]$ such that
$$
c_0^{-1}\frac{\rho^{2-s_h}-1}{(\frac{1}{\sqrt{p}})^{2-s_h}-1}\leq \widetilde{R}_0(x_0,x)\leq c_0\frac{\rho^{2-s_h}-1}{(\frac{1}{\sqrt{p}})^{2-s_h}-1}
$$
\end{proposition}

In fact, Corollary \ref{dc1} implies
$$
\widetilde{R}_0(x_0,x)\geq 2(1-\frac{1}{\nu})[1+\frac{1}{\nu p}+...+\frac{1}{(\nu p)^{d_h(x_0,x)-2}}],
$$
$$
\widetilde{R}_0(x_0,x)\leq 2[1+\frac{1}{\nu p}+...+\frac{1}{(\nu p)^{d_h(x_0,x)-1}}].
$$
In order to prove the proposition above, it remains to use the formula for the sum of a geometric progression with the ratio $\nu p<1$ and the Definition \ref{defsh}.

The following uniform in $s_h\in [1,2]$ estimate on $N_0(V)$ is a direct consequence of Theorem \ref{t1vv}, Proposition \ref{prunif} and Remark 2 after Theorem \ref{t1}.
\begin{theorem}
There exists a constant $C$ which is independent of $s_h\in [1,2]$ such that
\begin{equation}\label{1z1}
N_0(V)\leq 1+\#\{x:~V(x)\geq 1\}+C_1\sum_{x:V(x)<1}V(x)\frac{[\rho(x_0,x)]^{2-s_h}-1}{(\frac{1}{\sqrt{p}})^{2-s_h}-1}, \quad s_h<2;
\end{equation}
$$
N_0(V)\leq 1+\#\{x:~V(x)\geq 1\}+C_1\sum_{x:V(x)<1}V(x)\frac{\ln \rho(x,y)}{\ln\frac{1}{\sqrt p}}, \quad s_h=2.
$$
\end{theorem}
Similar arguments lead to the following statement.
\begin{theorem}
Estimate (\ref{1z1})  with a constant $C_1$ independent of $s_h$ remains valid when $s_h\in(2,3)$.
\end{theorem}

\end{document}